\documentclass[11pt,a4paper]{article}
\pdfoutput=1
\usepackage{jheppub}
\usepackage{graphicx}
\usepackage{epsfig}
\usepackage{amsmath, amssymb}
\usepackage{dcolumn}% Align table columns on decimal point
\usepackage{bm}% bold math
\usepackage{amsfonts}
\usepackage[toc,page]{appendix} %titletoc
\usepackage{color}

\usepackage{subfigure}
\usepackage{amssymb}
\usepackage{epstopdf}
\usepackage{enumerate}

\input epsf
\allowdisplaybreaks

\newcommand{\be}{\begin{equation}}
\newcommand{\ee}{\end{equation}}

\newcommand{\nn}{\nonumber \\}

\newcommand{\mn}{{\mu\nu}}
\newcommand{\ab}{{\alpha\beta}}

\newcommand{\hV}{\hspace{.5cm}}

\newcommand{\lb}{\left\lbrace}
\newcommand{\rb}{\right\rbrace}

\begin{document}

\title{Lattice implementation of Abelian gauge theories with 
Chern-Simons number and an axion field}

\author{Daniel G. Figueroa$^1$\,,}
\affiliation{$^1$CERN Theory Department, CH-1211 Geneve 23, Switzerland}
\emailAdd{daniel.figueroa@cern.ch}

\author{Mikhail Shaposhnikov$^2$}
\affiliation{$^2$Laboratory of Particle Physics and Cosmology Institute of  Physics, \'Ecole Polytechnique F\'ed\'erale de Lausanne, CH-1015 Lausanne, Switzerland}
\emailAdd{mikhail.shaposhnikov@epfl.ch}

\date{\today}

\abstract{Real time evolution of classical gauge fields is relevant for a number of applications in  particle physics and cosmology, ranging from the early Universe to dynamics of quark-gluon plasma. We present a lattice formulation of the interaction between a $shift$-symmetric field and some $U(1)$ gauge sector, $a(x)F_\mn\tilde F^\mn$, reproducing the continuum limit to order $\mathcal{O}(dx_\mu^2)$ and obeying the following properties: (i) the system is gauge invariant and (ii) shift symmetry is exact on the lattice.  For this end we construct a definition of the {\it topological number density} $Q = F_\mn\tilde F^\mn$ that admits a lattice total derivative representation $Q = \Delta_\mu^+ K^\mu$, reproducing to order $\mathcal{O}(dx_\mu^2)$ the continuum expression $Q = \partial_\mu K^\mu \propto \vec E \cdot \vec B$. If we consider a homogeneous field $a(x) = a(t)$, the system can be mapped into an Abelian gauge theory with Hamiltonian containing a Chern-Simons term for the gauge fields. This allow us to study in an accompanying paper the real time dynamics of fermion number non-conservation (or chirality breaking) in Abelian gauge theories at finite temperature.  When $a(x) = a(\vec x,t)$ is inhomogeneous, the set of lattice equations of motion do not admit however a simple explicit local solution (while preserving an $\mathcal{O}(dx_\mu^2)$ accuracy). We discuss an iterative scheme allowing to overcome this difficulty.}

 %\pacs{To be done}

\maketitle

%=====================================================================
%=====================================================================
%=====================================================================

\section{Introduction}
\label{sec:Intro}

Real-time evolution of classical fields has many applications in different areas of high energy physics and cosmology. These include the creation and evolution of topological defects in the early Universe~\cite{Grigoriev:1988bd,Vincent:1997cx,Rajantie:1999mp,Davis:2000kv,Hindmarsh:2001vp,Hindmarsh:2000kd,Bevis:2006mj,Figueroa:2012kw,Daverio:2015nva,Hindmarsh:2017qff}, non-perturbative investigations of hot sphaleron transitions related to fermion number non-conservation in the electroweak theory~\cite{Grigoriev:1989je,Ambjorn:1990pu,Ambjorn:1990wn,Moore:1998swa,Moore:1999fs,GarciaBellido:1999sv,Moore:2000mx,GarciaBellido:2003wd,Tranberg:2003gi,Tranberg:2006dg,DOnofrio:2014rug}, %(for an ultimate reference see~\cite{DOnofrio:2014rug} and references therein), 
the analysis of inflationary preheating~\cite{Khlebnikov:1996mc,Prokopec:1996rr,Felder:2000hj,Rajantie:2000nj,GarciaBellido:2002aj,DiazGil:2007dy,DiazGil:2008tf,Figueroa:2015rqa,Enqvist:2015sua,Figueroa:2016wxr}, the generation of cosmological perturbations~\cite{Bassett:1998wg,Bassett:1999mt,Bassett:1999ta,Finelli:2000ya,Chambers:2007se,Bond:2009xx} and gravitational waves~\cite{Khlebnikov:1997di,Easther:2006gt,GarciaBellido:2007dg,GarciaBellido:2007af,Dufaux:2007pt,Dufaux:2008dn,Figueroa:2011ye,Bethke:2013aba,Bethke:2013vca,Figueroa:2016ojl} during preheating, and different aspects of quark-gluon plasma, see e.g.~\cite{Attems:2016ged} and references therein. The classical approximation to quantum dynamics should work when the relevant distance scales of the problem exceed considerably the typical quantum distances, at finite temperatures $T$ given by $1/T$ (distance between particles) and $1/(gT)$ (Debye screening length, where $g$ is the gauge coupling). 

The numerical procedures for Abelian or non-Abelian scalar-gauge theories, such as the Standard Model is well developed, see e.g.~\cite{Ambjorn:1990pu} for first studies in 3+1 dimensions. In the simplest realisation it consists in the following steps (for inclusion in numerical simulations of hard thermal loops and Langevin-like dynamics see \cite{Bodeker:1999gx}). One uses the standard lattice formulation of the gauge theory action  with exact lattice gauge invariance and Minkowski signature of the metric.  The variation of the action with respect to the gauge fields (living at the links of the lattice) and scalar fields (living at the lattice sites) produce a system of equations in which the dynamical variables at time slice $t_n$ are expressed via those at two preceding times $t_{n-1}$ and $t_{n-2}$. So, giving an initial condition at $t_0$ and $t_1$ (the Cauchy problem)  allows to follow the evolution of the system and address all sorts of questions one is interested in. The initial conditions are chosen depending on the physical system under consideration: for sphaleron transitions they are taken from an equilibrium ensemble at some temperature, for preheating from knowing the spectrum of initial fluctuations of the fields after inflation, etc. 

It is very important that the procedure discussed above is gauge invariant. The variation of the action with respect to the zero component of the gauge field gives the lattice Gauss constraint, which is {\em exactly} conserved during the (lattice) time evolution. The gauge invariance insures the stability of numerics and keeps an (approximate) energy conservation; features that are necessary for a continuum interpretations of the results. It is an empirical fact that the formulations that are not gauge invariant in discrete space-time (and only invariant when the lattice spacing and time step go to zero) are plagued with different non-physical numerical instabilities. 

For a number of applications the simplest gauge-scalar actions should be extended by an addition of the pieces that contain the so called topological term or $Pontryagin$ density, $Q = F_\mn \tilde{F}^\mn$, where the dual of the field strength is defined as usual by $\tilde{F}^\mn \equiv {1\over 2}\epsilon^{\mn\ab}F_{\ab}$, with $\epsilon^{\mn\ab}$ being the completely anti-symmetric tensor in four dimensions, with $\epsilon^{0123} \equiv 1$.  The most interesting examples contain an axion field coupled linearly to $Q$ as $a(x)F_\mn\tilde F^\mn$, and the theories with non-zero chemical potential $\mu$ for chiral fermions, leading to an effective bosonic Hamiltonian containing the Chern-Simons term for the gauge fields $\mu n_{\rm cs}$, with $n_{\rm cs} \propto \int d^4x\,Q$. 

To investigate the time evolution of these systems one is faced with the following problem. The realisation of  $Q$ on the lattice should be done in such a way that the continuum topological properties of $Q$ hold: the integral of $Q$ over the volume of space-time can be expressed via an integral over the boundary.  To put it in other words, in continuum one can write 
\begin{equation}
Q=\partial_\mu K^\mu~,
\label{tc}
\end{equation}
where $K^\mu$ is the Chern-Simons current. The lattice analogue of this relation is 
\begin{equation}
Q=\Delta_\mu^+ K^\mu,
\label{ll}
\end{equation}
where $\Delta_\mu^+$ is the lattice difference in positive direction of $\mu$ axis (more details are provided in Sections~\ref{sec:LatticeBasics}, \ref{sec:LatticeFormulation} and \ref{sec:CSnum}). If the lattice (gauge-invariant) definition of $Q$ does not satisfy this property, we will get unwanted lattice artifacts and the continuum extrapolation would be difficult.  There are several interesting quantities which are very sensitive to the topological property (\ref{tc}). For example, it is Eq.~(\ref{tc}) which makes the axion mass $m_a$ to be zero in all orders of perturbation theory ($m_a\neq 0$ being a non-perturbative phenomenon). The extraction of the non-Abelian sphaleron rate in the symmetric phase of the electroweak theory from diffusion of Chern-Simons number requires a careful construction of the lattice version of $Q$ in which the property (\ref{tc}) is still approximate, but as precise as possible  \cite{Moore:1999fs}.

The aim of the present paper is to set up a lattice gauge invariant formulation for real time simulations of {\em Abelian} gauge theories, respecting the topological property Eq.~(\ref{ll}) of $Q$ {\em exactly}. As for non-Abelian theories, the mission seems to be impossible due to well known difficulties of defining $Q$ obeying the property (\ref{ll}) on the lattice \cite{Luscher:1981zq,Woit:1983tq}.  The applications of our formulation can include, for example, the study of the late stage of inflation and preheating in axion-inflation models~\cite{Freese:1990rb,Sorbo:2011rz,Cook:2011hg,Barnaby:2011qe,Linde:2012bt,Pajer:2013fsa,Adshead:2015pva,Adshead:2016iae}, the clarification of the role of Standard Model hypercharge group U(1) in baryogenesis and in magnetic field generation \cite{Joyce:1997uy,Giovannini:1997eg,Long:2013tha,Long:2016uez}, the modeling the different aspects of chiral magnetic effects~\cite{Vilenkin:1980fu,Joyce:1997uy,Alekseev:1998ds,Fukushima:2008xe}, or the study of the problem of chiral fermionic charge evolution in high temperature electrodynamics~\cite{Boyarsky:2011uy, Boyarsky:2012ex}. An accompanying paper~\cite{Figueroa:2017tbf} is devoted to the last subject. 

This paper is organised as follows. In Sect.~\ref{sec:AbelianTheoryContinuum} we review the classical equations of motion in the continuum of an Abelian gauge theory with an axion field. We also discuss how to map that system into an Abelian gauge theory with chemical potential. In Sect.~\ref{sec:LatticeBasics} we review the essence of the non-compact lattice formulation of an Abelian gauge theory, so that we set notation and conventions for the following sections. In Sect.~\ref{sec:LatticeFormulation} we build a lattice implementation of an axionic-interaction $a(x)\tilde{F}_\mn F^\mn$, deriving step by step the necessary ingredients to achieve a formulation consistent with the (lattice version) of the Bianchi identities, and solvable by an iterative scheme of evolution. We first consider in Sect.~\ref{subsec:LatticeHomAxion} the case of a homogeneous axion $a(x) = a(t)$, and later generalize to a fully inhomogeneous axion $a(x) = a(t,{\bf x})$ in Sect.~\ref{subsec:LatticeInHomAxion}. In Sect.~\ref{sec:CSnum} we finally discuss the lattice formulation of the Chern-Simons number $n_{\rm cs} \propto \int d^4x\,Q$ based on the lattice version(s) of $Q = \tilde{F}_\mn F^\mn$ developed in Sect.~\ref{sec:LatticeFormulation}. We put special care in the need to achieve a lattice formulation that admits a total derivative representation for $Q$ as in Eq.~(\ref{ll}). In Sect.~\ref{sec:Conclusions} we summarize our results and discuss some of the potential applications in particle physics and cosmology.

\section{Abelian gauge theory with an axion. Theory in the continuum}
\label{sec:AbelianTheoryContinuum}

Let us begin by considering the action of an Abelian gauge theory in flat space-time\footnote{We choose a metric signature $\eta_\mn = \eta^\mn = (-,+,+,+)$.}, in the presence of an axion-type field $a(x)$ linearly coupled to the $Pontryagin$ density $F_\mn \tilde{F}^\mn$ of a U(1) gauge field,
\begin{eqnarray}\label{eq:ActionContinuum}
S &=& -\int d^4x\left(\mathcal{L}_{\varphi} + {1\over4e^2}F_\mn F^\mn - {1\over 2c_s^2}(\partial_0 a)^2 + {1\over 2}(\partial_i a)(\partial_i a) - {1\over (4\pi)^2}{a\over M}F_\mn \tilde{F}^\mn\right)\nn
%&=& \int d^4x\left(\mathcal{L}_{\varphi} + \sum_i {F_{0i}^2\over 2e^2} - \sum_{i\neq j} {F_{ij}^2\over 4e^2} + {(\partial_o a)^2\over2 c_s^2} - \sum_i {(\partial_i a)^2\over2} + {1\over 8\pi^2}{a\over M}\sum_{i,j,k}F_{0i}\epsilon_{ijk}F_{jk}\right)\nn
&=& \int d^4x\left(-\mathcal{L}_{\varphi} + {1\over 2e^2}\left({\vec E}^2- {\vec B}^2\right) + {1\over2 c_s^2}{\dot a}^2 - {1\over2}|\nabla a|^2 + {1\over 4\pi^2}{a\over M}\vec{E}\vec{B}\right)\,,
\end{eqnarray}
%where summation over 4-dimensional repeated indexes is implicitly assumed from now on. %, whereas summation over 3-dimensional repeated indexes was made explicit. 
We consider a 'Higgs' sector as $\mathcal{L}_{\varphi} = (D_0\varphi)^*(D_0\varphi) - (\vec{D}\varphi)^*(\vec{D}\varphi) + V(\varphi^*\varphi)$, with $\varphi = (\varphi_1 + i\varphi_2)/\sqrt{2}$ a $U(1)$ charged field [with $\varphi_i \in \Re$], $A_\mu = (\phi,\vec{A})$ the gauge field, $e^2$ the gauge coupling strength, and $D_\mu \equiv \partial_\mu - iA_\mu$ the covariant derivative. The field strength and its dual are defined as usual by $F_\mn \equiv \partial_{\mu}A_{\nu}-\partial_{\nu}A_{\mu}$ and $\tilde{F}^\mn \equiv {1\over 2}\epsilon^{\mn\ab}F_{\ab}$. %, where $\epsilon^{\mn\ab}$ is the completely anti-symmetric tensor in four dimensions, with $\epsilon^{0123} \equiv 1$. 
Given our choice of $A_\mu \equiv (\phi,\vec A)$, we define the electric and magnetic fields as $E^i = E_i = \dot{A}_i - \partial_i\phi$ and $B^i = B_i = \epsilon_{ijk}\partial_jA_k$. This leads to the relations $E^i \equiv F_{0i} = - F^{0i}$, $B^i \equiv {1\over2}\epsilon^{ijk}F_{jk}$, so that $F_{\mn} = (\delta_{\mu0}\delta_{\nu i}-\delta_{\mu i}\delta_{\nu 0})E^i + (\delta_{\mu i}\delta_{\nu j}-\delta_{\mu j}\delta_{\nu i})\epsilon_{ijk}B^k$. %, with $\epsilon_{ijk}$ the completely anti-symmetric tensor in three dimensions, with $\epsilon_{123} \equiv 1$. 
Given our metric signature, we obtain $F_{\mn}\tilde{F}^{\mn} = +4\vec E\vec B$, and arrive at the final vectorial expressions of Eq.~(\ref{eq:ActionContinuum}). 

Lagrangian Eq.~(\ref{eq:ActionContinuum}) describes a system of scalar electro-dynamics in the presence of an axion-like field $a(x)$, with $M$ some mass scale undetermined at this point. Note that we maintain explicitly the speed of propagation of the axion $c_s^2$ as a free parameter, as this will be convenient for us later on. Action Eq.~(\ref{eq:ActionContinuum}) is invariant under the transformations $\varphi(x) \rightarrow e^{+i\beta(x)}\varphi(x)$, $A_\mu(x) \rightarrow A_\mu(x) +\partial_\mu\beta(x)$, with $\beta(x) \in \Re$ and $e^{i\beta(x)} \in U(1)$. Varying the action, one obtains the equations of motion (EOM)
\begin{eqnarray}\label{eq:EOM1}
D_\mu D^\mu \varphi &=& V,_{\varphi^*}\,,\\
\label{eq:EOM2}
\partial_\nu F^\mn - {e^2\over 4\pi^2}{a\over M}\partial_\nu \tilde{F}^\mn &=& e^2j^\mu + {e^2\over 4\pi^2}{\partial_\nu a\over M}\tilde{F}^\mn\,,\\
\label{eq:EOM3}
\partial_0\partial_0 a - c_s^{2}\partial_i\partial_i a &=& {c_s^{2}\over 4\pi^2 M}{F}_\mn\tilde{F}^\mn\,,
\end{eqnarray}
where the (unit-charge) current is defined as $j^\mu = 2{\rm Im}\lbrace\varphi^* D^\mu\varphi\rbrace$, so that 
\begin{equation}
j_\mu = (\rho,\vec J) \equiv (2{\rm Im}\lbrace\phi^*\dot\phi\rbrace,2{\rm Im}\lbrace\phi^*\vec D\phi\rbrace)
\end{equation}
Eqs.~(\ref{eq:EOM1})-(\ref{eq:EOM3}) can be rewritten in a vectorial form as
\begin{eqnarray}\label{eq:EOMvector1}
D_oD_o\varphi - \vec D \vec D\varphi &=& -V_{,|\varphi|^2}\varphi\,,\\
\label{eq:EOMvector2}
\dot{\vec E} + \vec\nabla \times \vec B + {e^2\over 4\pi^2}{a\over M}(\dot{\vec B} - \vec\nabla \times \vec E) &=& e^2\vec{J} - {e^2\over 4\pi^2 M}{\dot a}\vec B + {e^2\over 4\pi^2 M}{\vec\nabla a}\times \vec E\,,\\
\label{eq:EOMvector3}
\vec\nabla \vec E + {e^2\over 4\pi^2}{a\over M}\vec\nabla\vec B &=& e^2\rho - {e^2\over 4\pi^2 M}{\vec\nabla a}\cdot \vec B\,,\\
\label{eq:EOMvector4}
\ddot a - c_s^2{\vec\nabla}^2a &=& {c_s^2\over 4\pi^2 M}\vec E \cdot \vec B\,,
\end{eqnarray}
with Eq.~(\ref{eq:EOMvector3}) representing the Gauss constraint in the presence of an axion.

As mentioned in Section~\ref{sec:Intro}, it is well known that the Pontryagin density represents a topological term, as it can be written as a total derivative $F_\mn\tilde{F}^\mn = \partial_\mu K^\mu$. %Hence such a term is expected to not contribute into the fields' EOM. 
This is reflected by the Bianchi identities, i.e.~the term $\partial_\nu \tilde{F}^{\mn} = 0$ in Eq.~(\ref{eq:EOM2}), or equivalently its vectorial counterparts $(\dot{\vec B} - \vec \nabla \times E) = 0$ in Eq.~(\ref{eq:EOMvector2}), and $\vec \nabla \vec B = 0$ in Eq.~(\ref{eq:EOMvector3}). Those terms simply represent vanishing contributions in the EOM, so it is customary to remove them. It is nonetheless convenient for us to keep such terms in the EOM, despite their null contribution. The reason for this will become clear, however, only in Sec.~\ref{sec:LatticeFormulation}, after we introduce the lattice discretization scheme(s) for the action Eq.~(\ref{eq:ActionContinuum}). 

For the time being, let us simply note now that due to the topological nature of the $F_\mn\tilde{F}^\mn$ operator, action Eq.~(\ref{eq:ActionContinuum}) is also ('topologically') invariant under $a(x) \rightarrow a(x) + C$, with $C$ an arbitrary constant. This is reflected in the fact that the linear coupling of $a(x)$ to the Pontryagin density $\propto \int d^4x \,a(x)\,F_\mn\tilde F^\mn$, represents in reality a derivative coupling: thanks to the total derivative nature of $F_\mn\tilde{F}^\mn = \partial_\mu K^\mu$, after integration by parts, we obtain $\propto \int d^4x \,K^\mu\partial_\mu a(x)$. Once the Bianchi identities are considered, the terms proportional to $a(x)$ in Eq.~(\ref{eq:EOM2}) [equivalently in Eqs.~(\ref{eq:EOMvector2}), (\ref{eq:EOMvector3})] disappear, as it must, for a derivative coupling. As we will see later in Sec.~\ref{sec:LatticeFormulation}, the lattice equivalent of the Bianchi identities, and hence the topological nature of the lattice equivalent of $F_\mn\tilde F^\mn$, depend crucially on the lattice discretization scheme. The lattice equivalent terms to $\propto \vec\nabla\vec B$, $\propto (\dot{\vec B} - \vec \nabla \times E)$ in the discrete EOM, are actually not granted to vanish by default. Achieving such a goal will represent, in fact, a guiding principle towards the construction of a correct lattice formulation of $F_\mn\tilde{F}^\mn$, admitting a total derivative representation on the lattice.

Let us remark that the $shift$ symmetry enjoyed by $a(x)$, together with the linear nature of its coupling to the Pontryagin density, is at the heart of the original introduction of the axion as a solution for the strong CP problem\footnote{Of course the QCD axion is rather coupled to Tr\,$\tilde G_\mn G^\mn$, where $G_\mn$ is the gluon field strength.}. Once $a(x)$ is considered as a dynamical field, the interaction $a(x)F_\mn\tilde{F}^\mn$ in Eq.~(\ref{eq:ActionContinuum}) leads naturally to extra contributions in the EOM. Assuming a fully space-time dependent axion field, naturally leads to a contribution in Eq.~(\ref{eq:EOM2}) as $\propto \tilde{F}^{\mn}\partial_\nu a$, or equivalently by the vectorial counterparts $\propto (-{\dot a}\vec B + {\vec\nabla a}\times \vec E)$ and $\propto {\vec\nabla a}\cdot \vec B$ in Eq.~(\ref{eq:EOMvector2}) and Eq.~(\ref{eq:EOMvector3}), respectively. Besides, the Pontryagin density $F_\mn\tilde{F}^\mn$ acts as a source for $a(x)$ in the $rhs$ of Eq.~(\ref{eq:EOMvector4}). The promotion of $a(x)$ into a dynamical field can therefore affect notably the dynamics of the system with respect to standard scalar electrodynamics described by Eq.~(\ref{eq:ActionContinuum}) with $a(x) = 0$.

\subsection{Mimicking a chemical potential}

Interestingly, through an adequate interpretation of field variables and parameters, Eq.~(\ref{eq:ActionContinuum}) can be mapped into the description of a gauge theory with a chemical potential $\mu$ for chiral fermions (for more details see \cite{Figueroa:2017tbf}). Starting from Eq.~(\ref{eq:ActionContinuum}) will allow us, through a formal trick, to bring up a Lagrangian formulation into this problem. In order to see this, let us begin by demanding that $a(x) = a(t)$ is a spatially homogeneous field, so that
\begin{equation}
\vec \nabla a = 0\,.
\end{equation}
We then introduce the following convenient 'dimensionally reduced' variables
\begin{equation}\label{eq:dimensionallyReducedVariables}
a \equiv \alpha M\,,~~\dot a \equiv \mu M\,,
\end{equation}
so that Eq.~(\ref{eq:ActionContinuum}) is reduced to
\begin{eqnarray}\label{eq:ActionContinuumII}
S &=& \int d^4x\left(\mathcal{L}_{\varphi} + {1\over 4e^2}F_\mn F^\mn + M^2{{\dot\alpha}^2\over 2c_s^2} + {\alpha\over (4\pi)^2}F_\mn \tilde{F}^\mn\right) \nn
&=& \int d^4x\left\lbrace |D_0\varphi|^2 - (\vec{D}\varphi)^*(\vec{D}\varphi) + V(\varphi^*\varphi) + {1\over 2e^2}\left({\vec E}^2- {\vec B}^2\right)\right\rbrace \nn
&+& \lim_{V\rightarrow\infty}\left\lbrace \int dt \,{{\dot\alpha}^2\over 2c_s^2}\int_V M^2d^3x + \int dt \,{\alpha\over 4\pi^2} \int_V d^3x \vec{E}\vec{B}\right\rbrace\,.
\end{eqnarray}
As we will see next, the requisite to describe a gauge theory at high temperatures and in the presence of a chemical potential, will fix the mass scale $M$ and parameter $c_s^2$, see Eq.~(\ref{eq:cs2}). We will describe first, however, the new dynamical equations that follow from minimizing the new re-written action. 

Varying Eq.~(\ref{eq:ActionContinuumII}) we obtain the equations of motion of the system, which we write directly in a vectorial form as
\begin{eqnarray}\label{eq:EOMChPot1}
D_oD_o\varphi - D_jD_j\varphi &=& -V_{,|\varphi|^2}\varphi\,,\\
\label{eq:EOMChPot2}
\dot{\vec E} + \vec\nabla \times \vec B &=& e^2\vec{J} - {e^2\over 4\pi^2}{\mu}\vec B - {e^2\over 4\pi^2}\alpha(\dot{\vec B} - \vec\nabla \times \vec E)\\
\label{eq:EOMChPot3}
\vec\nabla \vec E &=& e^2\rho - {e^2\over 4\pi^2}\alpha\vec\nabla\vec B ~~{\rm (Gauss\,\,Constraint)}\,,\\
\label{eq:EOMChPot4}
\dot \mu &=& {c_s^2\over 4\pi^2}{1\over M^2}\lim_{V\rightarrow\infty} {1\over V}{\int_V d^3x ~\vec E \cdot \vec B}\,.
\end{eqnarray}
Once again, let us note that the terms $\propto \alpha(\dot{\vec B} - \vec \nabla \times E)$ in Eq.~(\ref{eq:EOMvector2}) and $\propto \alpha\vec \nabla \vec B$, which vanish in the continuum, are only maintained in the above equations for later convenience when discretising the system in Sect.~\ref{sec:LatticeFormulation}.

We can now fix the mass scale $M$ and the parameter $c_s^2$ to appropriate values, so that the set of Eqs.~(\ref{eq:EOMChPot1})-(\ref{eq:EOMChPot4}) properly describe an Abelian gauge theory with the chemical potential $\mu$ for chiral fermionic charge. In particular, the EOM of a chemical potential follows from anomaly equation \cite{Adler:1969gk,Bell:1969ts} and in our case has the form~\cite{Boyarsky:2012ex}
\begin{equation}
\dot \mu = {3\over \pi^2}{1\over T^2} {1\over V}{\int_{_{V\rightarrow\infty}} \hspace*{-0.65cm} d^3x \,\,\vec E \cdot \vec B}\,.
\end{equation}
In light of Eq.~(\ref{eq:EOMChPot4}), we must identify the mass scale with the temperature of the system, 
\begin{equation}\label{eq:Mtemp}
M = T~ ({\rm  temperature})\,,
\end{equation}
and fix the dimensionless parameter to
\begin{equation}\label{eq:cs2}
c_s^2 = 12\,.
\end{equation}
Let us note that in the original action Eq.~(\ref{eq:ActionContinuum}), $c_s^2$ represents the speed of propagation of the axion field. However, as we considering the axion now as a homogeneous field $a(x) = a(t)$, $c_s^2$ represents simply a parameter in the theory. One should not conclude therefore that the value given by Eq.~(\ref{eq:cs2}) represents a super-luminal axion speed of propagation, as in the chemical potential context where such value is determined, $c_s^2$ does not represent, in first place, a propagation speed. Eqs.~(\ref{eq:EOMChPot1})-(\ref{eq:EOMChPot4}) together with Eqs.~(\ref{eq:Mtemp})-(\ref{eq:cs2}), describe appropriately an Abelian gauge with chemical potential $\mu$.

\section{Part I. Lattice formulation of Abelian gauge theories to order $\mathcal{O}(dx_\mu^2)$, Scalar-Electrodynamics}
\label{sec:LatticeBasics}

In this section we briefly summarize the basics of the lattice formulation of a gauge theory. We focus, for convenience, in the non-compact formulation of an Abelian gauge theory (for related discussion see, e.g. \cite{Kajantie:1998rz}). We just intent to set notation and basic concepts, which we will be used later on when discussing the lattice formulation of an Abelian gauge theory with an axion. A reader already familiar with lattice gauge invariant techniques can skip this section and jump directly into Sect.~\ref{sec:LatticeFormulation}.

\subsection{Non-compact formulation of Scalar-Electrodynamics}

Let us, first of all, set some notation. We will not consider summation over repeated indices, as this can lead to confusion. A lattice point $n = (n_o,\vec n) = (n_o,n_1,n_2,n_3)$ displaced in the $\mu-$direction by one unit lattice spacing, $n + \hat\mu$, will be often referred simply as $n+\mu$ or by $+\mu$. For example, $\varphi_{+\mu} \equiv \varphi(n+\hat\mu)$. Components of gauge fields live in between lattice sites in the direction of the component, so $A_{\mu} \equiv A_{\mu}(n+{1\over2}\hat\mu)$, $A_{\mu,+\nu} \equiv A_{\mu}(n+{1\over2}\hat\mu + \hat\nu)$, etc. For simplicity of the notation, we will refer to both the lattice spacing $\Delta x$ and the time step $\Delta t$, simply as $dx$, so if we write e.g.~$+\hat \mu dx$, this should be interpreted as a time step advancement $+\Delta t$ if $\mu = 0$, or as a unit displacement $+\Delta x$ in a given spatial direction $\mu = 1,2$ or $3$. We will loosely speak of the lattice spacing order $\mathcal{O}(dx)$, independently or whether we are referring to $\mathcal{O}(\Delta t)$ or $\mathcal{O}(\Delta x)$.

We define a lattice {\it link}, as usual, like $U_{\mu} \equiv U_\mu(n+{1\over 2}\hat\mu) \equiv e^{-i\int_{x(n)}^{x(n+\hat\mu)} A_\mu(x')dx'^\mu}$ $\simeq e^{-idx^\mu A_\mu(n+{1\over 2}\hat\mu)}$. We also define $U_{-\mu} \equiv U_{\mu,-\mu}^* \equiv U_\mu^*(n-{1\over 2}\hat\mu) \simeq e^{+idx^\mu A_\mu(n-{1\over 2}\hat\mu)}$. Forward (+) and backward (-), ordinary and covariant derivatives, are defined in the lattice by
\begin{eqnarray}
\begin{array}{rclr}
\Delta_\mu^\pm\phi \equiv {\pm 1\over dx}(\phi_{\pm \mu} - \phi) & \rightarrow & (\partial_{\mu}\varphi)(x) + O(dx^2) &\,,~ x \equiv  (n\pm{1\over 2}\hat\mu)dx\,,\vspace*{0.2cm}\\
(D_\mu^\pm\varphi) \equiv {\pm 1\over dx}(U_{\pm \mu}\varphi_{\pm \mu}-\varphi) & \rightarrow & (D_{\mu}\varphi)(x)\left(1 - {1\over 2}idxA_\mu(x)\right) &,~ x \equiv (n\pm{1\over 2}\hat\mu)dx\,,
\end{array}
\end{eqnarray}
where we have indicated the order in the lattice spacing to which one recovers the continuum limit, as well as the natural space-time location in the continuum the derivatives live. A lattice gauge transformation under $U(1)$ corresponds to 
\begin{eqnarray}
\varphi(n) ~~\longrightarrow~~ e^{+i\beta(n)}\varphi(n)\,,~~~~~~ A_{\mu}(n+{1\over 2}\hat\mu) ~~\longrightarrow~~ A_{\mu}(n+{1\over 2}\hat\mu) \,+\, \Delta_{\mu}^+\beta(n+{1\over 2}\hat\mu),
\end{eqnarray}
with $\beta$ an arbitrary function, so that the links and covariant derivatives transform as
\begin{eqnarray}
U_{\pm\mu,n} ~~\longrightarrow~~ e^{i\beta}\,U_{\pm\mu,n}\,e^{-i\beta_{\pm\mu}}\,,~~~~~~ D_\mu^\pm\varphi ~~\longrightarrow~~ e^{i\beta}\,D_\mu^\pm\varphi\,.\hspace*{4cm}
\end{eqnarray}
We will use these transformation rules to build a gauge invariant lattice action in the following. In this section we ignore the axion field, so for the time being we just consider a lattice action for scalar-electrodynamics only. Using a non-compact formulation, we can write
\begin{eqnarray}\label{eq:ActionLAH}
S_{\rm AH}^{L} &=& \Delta t \Delta x^3\sum_{\vec n,t} \lb %\frac{1}{2}\Delta_o^+\phi - \frac{1}{2}\sum_j\Delta_j^+\phi + 
(D_o^{+}\varphi)^\dag(D_o^{+}\varphi) - \sum_j(D_j^{+}\varphi)^\dag(D_j^{+}\varphi) - V(\varphi\varphi^*,\phi) \right. \nn 
&& \hspace*{2.0cm}\left. +~ \frac{1}{2e^2}\sum_{j}\left(\Delta_o^+A_i-\Delta_i^+A_o\right)^2  - \frac{1}{4e^2}\sum_{i,j}(\Delta_i^+A_j-\Delta_j^+A_i)^2 \rb\,,
\end{eqnarray}
from where the lattice gauge invariance (based on the transformations defined above) is rather explicit. We will refer to Eq.~(\ref{eq:ActionLAH}) as the $Abelian-Higgs$ (AH) lattice action. As we will show in Section~\ref{subsec:ContinuumLimit}, this action reproduces to order $\mathcal{O}(dx^2)$ the continuum action Eq.~(\ref{eq:ActionContinuum}) in the absence of an axion field ($a(x) = 0$). Varying Eq.~(\ref{eq:ActionLAH}) with respect to the different fields, we obtain the lattice equivalent of the dynamical equations, which read (taking the Coulomb Gauge $A_o = 0 \leftrightarrow U_o = 1$)
\begin{eqnarray}\label{eq:EOMlattice1}
&& \Delta_o^-\Delta_o^+\varphi - \sum_iD_i^-D_i^+\varphi + V_{,\varphi^*} = 0\\
\label{eq:EOMlattice2}
&& \Delta_o^-\Delta_o^+(A_i) - \sum_{j}\left(\Delta_j^-\Delta_j^+(A_i) - \Delta_i^+ \Delta_j^-(A_j)\right) = 2e^2{\rm Im}\lbrace\varphi^*D_i\varphi\rbrace\\
\label{eq:EOMlattice3}
&& \sum_i \Delta_i^-\Delta_o^+(A_i) = 2e^2{\rm Im}\lbrace\varphi^*\Delta_{+o}\varphi\rbrace\hV(\rm Gauss\,\,Constraint)
\end{eqnarray} 

\subsection{Recovering the continuum limit to order $\mathcal{O}(dx^2)$}
\label{subsec:ContinuumLimit}

Let us note that, given a continuum action $S = \int dt\,d^3x\, \mathcal{L}_{C}$, with lagrangian given by the sum of various operators $\mathcal{L}_{C} = \sum_p \mathcal{O}_{C}^{(p)}$, we try to emulate the same physical system by defining a lattice action $S = \Delta t \Delta x^3 \sum_{\vec n,n_o}\mathcal{L}_{L}$, with the lattice lagrangian given by the sum of lattice operators $\mathcal{L}_{L} = \sum_p \mathcal{O}_{L}^{(p)}$. Each operator $\mathcal{O}_{L}^{(p)}$, when expanded around the lattice site where it is defined, reproduces the continuum operator $\mathcal{O}_{C}^{(p)}$ to some $n$-th order in the lattice spacing, $\mathcal{O}_{L}^{(p)} \rightarrow \mathcal{O}_{C}^{(p)} + \mathcal{O}(dx^n)$. For consistency we require that each and every operator $\mathcal{O}_{L}^{(p)}$ reproduces the corresponding continuum term $\mathcal{O}_{L}^{(p)}$ to the same $n$-th order. 

Let us inspect in detail action Eq.~(\ref{eq:ActionLAH}). Each term in the action reproduces the equivalent continuum terms to second order in $dx$, i.e.~$\mathcal{O}_{\rm L}^{(p)} \rightarrow \mathcal{O}_{\rm C}^{(p)} + \mathcal{O}(dx^2),~\forall p$. In order to verify this,  it is crucial to make a suitable interpretation of the lattice sites where each operator naturally 'lives'. For instance, let us consider a lattice operator mimicking the kinetic term of a scalar field, $\mathcal{O}_{L}^{(k)} = {1\over 2}(\Delta_\mu^+\phi)^2$. If we were to expand $\Delta_\mu^+\varphi$ around the position $x = n\cdot dx$, we would obtain $(\Delta_\mu^+\phi)(n) \rightarrow (\partial_\mu\phi)({\vec x}) + \mathcal{O}(dx)$. Consequently $\mathcal{O}_{L}^{(k)} \rightarrow \mathcal{O}_{C}^{(k)} + \mathcal{O}(dx)$, so the continuum limit is only reproduced to linear order. The natural site where $\mathcal{O}_{L}^{(k)}$ really lives is however, not $x = n dx$, but rather in $x_{\mu/2} \equiv (n+{1\over 2}\hat\mu) dx$, as it involves a finite difference of the field evaluated (with equal weight) at both $n$ and $n+\hat\mu$ lattice sites. If we expand $\mathcal{O}_{L}^{(k)}$ around ${x}_{\mu/2}$, then we obtain $(\Delta_\mu\phi)(n+{1\over 2}\hat\mu) \rightarrow (\partial_\mu\phi)({x}_{\mu/2}) + \mathcal{O}(dx^2)$. Therefore, in order to analyze the continuum limit of a lattice operator, one must first recognize which is the natural lattice site where it lives, and only then expand around this site. This will lead to the %(at first sight, apparently contradictory) 
fact that each operator $\mathcal{O}_{L}^{(p)}$ in the lattice lagrangian $\mathcal{L}_{L} = \sum_p \mathcal{O}_{L}^{(p)}$, may (possibly) live at different lattice sites. There is however no contradiction in this result, for as long as each operator $\mathcal{O}_{L}^{(p)}$ is interpreted as living in its natural lattice site, and reproducing the continuum limit to a common $n$-th order. Once this is imposed, the EOM obtained from varying a lattice action satisfying these two requisites, are guaranteed to reproduce the continuum EOM to the same order. Besides, each lattice EOM will live naturally in a well defined lattice site common to all the terms involved in the discrete equation, which determined the site around which to compute the continuum limit. Let us note, however, that the lattice site where different equations live, do not need to be the same.

Let us check all this with the terms of the action Eq.~(\ref{eq:ActionLAH}), as this will serve as a good training for our task in Sect.~\ref{sec:LatticeFormulation}, building an appropriate lattice operator for an axionic interaction. As mentioned before, each term in action Eq.~(\ref{eq:ActionLAH}) reproduces the continuum to second order in $dx$, i.e.~$\mathcal{O}_{\rm L}^{(p)} \rightarrow \mathcal{O}_{\rm C}^{(p)} + \mathcal{O}(dx^2),~\forall p$, for as long as each term is expanded around its natural lattice site. Let us check this term by term (recall there is no implicit sum over repeated indexes), denoting by $l$ the natural lattice site of each operator, and by $\vec x \equiv l dx$ the physical coordinate where the continuum limit is reproduced. A simple expansion leads to
\begin{eqnarray}
|D_\mu^+\varphi|^2(l)\big|_{{l \equiv n+{\hat\mu\over 2}}} &=& \left|e^{-idxA_\mu(l)}\varphi(l+\mu/2)-\varphi(l-\mu/2)\right|^2{1\over dx^2} \\
&\rightarrow & |D_\mu\varphi|^2(x)\left|1-{1\over 2}iA_\mu dx^\mu\right|^2 = |D_\mu\varphi|^2(x) + \mathcal{O}(dx^2)\,,\nn
&& \hspace*{4.7cm} \left[{x} \equiv ldx = (n + {1\over 2}\hat\mu)dx\right]\nn
\left(\Delta_\mu^+ A_\nu - \Delta_\nu^+ A_\mu\right)^2(l)\big|_{{l \equiv n + {\hat\mu\over 2} + {\hat\nu\over 2}}} &=& \left(A_\nu(l+{\hat\mu\over 2})-A_\nu(l-{\hat\mu\over 2})-A_\mu(l+{\hat\nu\over 2})+A_\mu(l-{\hat\nu\over 2})\right){1\over dx^2}  \nn
&\rightarrow& \left(\partial_\mu A_\nu - \partial_\nu A_\mu + \mathcal{O}(dx^2)\right)^2(x) = F_\mn^2(x) + \mathcal{O}(dx^2)\,,\nn
&& \hspace*{4.0cm} \left[{x} \equiv ldx = (n + {1\over 2}\hat\mu + {\hat\nu\over 2})dx\right]\nn
\end{eqnarray}
Let us turn our attention now into the lattice EOM Eqs.~(\ref{eq:EOMlattice1})-(\ref{eq:EOMlattice3}). Let us begin by the EOM of the charged scalar field, which involve the terms $V_{,\varphi^*}$ and $D_\mu^-D_\mu^+\varphi$.  We can expand each term around $l = n$, as the operator $D_\mu^+\varphi$ lives naturally at $n+{1\over 2}\hat\mu$, but $D_\mu^-(D_\mu^+\varphi)$ makes it live back to $l = (n+{1\over 2}\hat\mu) - {1\over 2}\hat\mu = n$,
\begin{eqnarray}
(D_\mu^-D_\mu^+\varphi)(l)\big|_{l \equiv n} &=& \left(e^{-idxA_\mu(l+\mu/2)}\varphi(l+\mu)+e^{-i(-dx)A_\mu(l-\mu/2)}\varphi(l-\mu)-2\varphi(l)\right){1\over dx^2} \nn
&\rightarrow& (D_\mu D_\mu\varphi)(l)\big|_{x \equiv ldx} + \mathcal{O}(dx^2)
\end{eqnarray}
Consistently, the term $V_{,\varphi^*}$ in Eq.~(\ref{eq:EOMlattice3}) [which reproduces the continuum to any order in $dx$], lives naturally at the same lattice site $l = n$. Therefore, when interpreting that the lattice operators involved in the discrete EOM of the charged scalar live at $l = n$, Eq.~(\ref{eq:EOMlattice3}) reproduces correctly the corresponding continuum Eq.~(\ref{eq:EOMChPot1}) up to $\mathcal{O}(dx^2)$ corrections. 

The EOM of the gauge fields contain more terms, but they all live consistently at the same lattice site, e.g.~at $l = n + {\hat j \over 2}$ for the dynamical equation Eq.~(\ref{eq:EOMlattice2}). This can be easily shown by expanding each term of this equation
\begin{eqnarray}
\Delta_o^-\Delta_o^+A_i(l)\big|_{l \equiv n + {\hat i\over2}} &=& \left(A_i(l+{\hat i\over2} + \hat 0) + A_i(l+{\hat i\over2} - \hat 0) -2A_i(l+{\hat i\over2})\right){1\over dt^2}\nn
&\rightarrow& (\ddot A_{i})(x)\big|_{x \equiv ndx + {\hat 0\over2}dt} + \mathcal{O}(dt^2)\\
\Delta_j^-\Delta_j^+A_i(l)\big|_{l \equiv n + {\hat i\over2}}  &=& \left(A_i(l+{\hat i\over2} + \hat j) + A_i(l+{\hat i\over2} - \hat j) -2A_i(l+{\hat i\over2})\right){1\over dx^2}\nn
&\rightarrow& (\partial_j^2 A_{i})(x)\big|_{x \equiv (n+{\hat i\over2})dx} + \mathcal{O}(dx^2)\\ 
\Delta_i^+ \Delta_j^-A_j(l)\big|_{l \equiv n + {\hat i\over2}} &=& \left(A_j(l+{\hat i\over2} + {\hat j\over2}) + A_j(l-{\hat i\over2} - {\hat j\over2})\right.\nn
&& \hspace*{1cm} \left. - A_j(l + {\hat i\over2} - {\hat j\over2}) - A_j(l - {\hat i\over2} + {\hat j\over2})\right){1\over dx^2}\nn
&\rightarrow& (\partial_i\partial_j A_{j})(x)\big|_{x \equiv (n+{\hat i\over2})dx} + \mathcal{O}(dx^2)\\
{\rm Im}\lbrace\varphi^*D_i^+\varphi\rbrace(l)\big|_{l \equiv n + {\hat i\over2}} &=& {\rm Im}\lbrace\varphi^*(l-{\hat i\over2})e^{-iA_i(l)dx}\varphi(l+\hat i /2)\rbrace/dx \nn
&\rightarrow& {\rm Im}\lbrace\varphi^*D_i\varphi\rbrace(x)\big|_{x \equiv (n+{\hat i\over2})dx} + \mathcal{O}(dx^2)
\end{eqnarray}
A similar analysis can be done expanding the terms of Eq.~(\ref{eq:EOMlattice1}) [Gauss constraint] around their natural lattice site $l = n + {1\over2}\hat 0$, leading to the same conclusion: when interpreting that the lattice operators involved in the discrete equations live at $l = n + {1\over2}\hat 0$ [Gauss constraint Eq.~(\ref{eq:EOMlattice1})] or $l = n + {1\over2}\hat i$ [Dynamical Eq.~(\ref{eq:EOMlattice2})], the discrete EOM of the gauge fields reproduce correctly the continuum equations Eqs.~(\ref{eq:EOMChPot2}), (\ref{eq:EOMChPot3}) [in the absence of an axion] up to $\mathcal{O}(dx^2)$ corrections.

\section{Part II. Lattice formulation of Abelian gauge theories to order $\mathcal{O}(dx_\mu^2)$, Axionic-coupling}
\label{sec:LatticeFormulation}

Let us now turn our discussion into the formulation of a proper lattice equivalent for the continuum interaction between the gauge fields and an axion, $S_{ac} \equiv {1\over(4\pi)^2}\int d^4x {a\over M}F_{\mn}\tilde{F}^{\mn}$. In this section we aim to a general formulation of an Abelian gauge theory with an axion field. For simplicity we will first start dealing with the case of a homogeneous axion $\alpha(x) = \alpha(t)$ in Sect.~\ref{subsec:LatticeHomAxion}. Our findings in Sect.~\ref{subsec:LatticeHomAxion} will be actually applicable as well to the case of a fully inhomogeneous axion $\alpha(x) = \alpha(t,{\bf x})$, but as the latter introduces further complications, we postpone the discussion about a fully spatially-dependent axion for Sect.~\ref{subsec:LatticeInHomAxion}. 

We will introduce the dimensionally reduced variables $\alpha \equiv {a\over M}$, $\mu \equiv \dot\alpha$ defined in Eq.~(\ref{eq:dimensionallyReducedVariables}), so we prevent this way having to drag the scale $M$ along our derivations. Given our choice of metric signature $(-,+,+,+)$ and gauge field representation $A_\mu \equiv (\phi,\vec A)$, we find $F_{\mn}\tilde{F}^{\mn} = +4\vec E\vec B$, so that we can write the above continuum  action as
\begin{equation}\label{eq:acContinuum}
S_{ac} \equiv {1\over 4\pi^2}\int d^4x\,\alpha\,\vec E\vec B\,.
\end{equation}
We will refer to the interaction described by Eq.~(\ref{eq:acContinuum}) as an $axionic-coupling$. Our main aim now is to formulate a latttice version of the continuum action Eq.~(\ref{eq:acContinuum}), from which to derive a discrete version of the EOM in the continuum Eqs.~(\ref{eq:EOMvector1})-(\ref{eq:EOMvector4}) [or Eqs.~(\ref{eq:EOMChPot1})-(\ref{eq:EOMChPot4}) in the case of a homogeneous axion mimicking a chemical potential]. In light of the EOM in the continuum, we can foresee three possible problems arising when formulating a lattice version of Eq.~(\ref{eq:acContinuum}):

\begin{enumerate}[i)]

\item The terms $\alpha(-\dot{\vec B}+\vec\nabla\times \vec E)$ and $\alpha\vec\nabla{\vec B}$ in Eqs.~(\ref{eq:EOMChPot2}), (\ref{eq:EOMChPot3}), vanish in the continuum thanks to the Bianchi identities $\partial_\nu\tilde{F}^{\mu\nu} = 0$, which are equivalent to $\dot{\vec B} = \vec\nabla\times \vec E$ and $\vec\nabla{\vec B} = 0$. The equivalent terms in the discrete EOM are however not granted to vanish, as this depends on the lattice representation of the electric and magnetic fields, and on the choice of lattice derivatives. It is therefore crucial that we find a lattice representation of $S_{ac}$ so that the equivalent discrete terms in the lattice EOM vanish identically (or at least to the same order in the lattice spacing to which the discrete EOM reproduce the continuum). In other words we seek a lattice formulation of $S_{ac}$ so that the lattice expression of $Q = \tilde{F}_{\mu\nu}\tilde{F}^{\mu\nu}$ is topological admitting a total derivative representation $Q = \Delta_\mu^+K^\mu$.

\item Assuming that a correct version of the Bianchi identities follows naturally from a given topological lattice formulation of ${F}_{\mu\nu}\tilde{F}^{\mu\nu}$, another problem may arise. The terms $\propto \mu \vec B$ and $\propto \vec E\vec B$ in Eqs.~(\ref{eq:EOMvector2}), (\ref{eq:EOMvector4}), indicate possible obstructions to achieving an explicit scheme to solve iteratively the set of lattice coupled equations reproducing the continuum Eqs.~(\ref{eq:EOMvector1})-(\ref{eq:EOMvector4}). Even though an implicit scheme for finite difference coupled equations can be solved by non-linear numerical methods (applied at every lattice site), this makes the continuum limit less transparent, and results typically in a computationally more expensive procedure (if not unfeasible). Therefore, achieving a simple explicit scheme for solving iteratively the set of lattice coupled equations that will mimic the continuum EOM, will be a strong requisite, unless we prove that such a scheme cannot be developed.

\item When considering a fully inhomogeneous axion-like field $a(x)$, terms proportional to spatial variations $\propto \nabla a(x)$ appear in the EOM. In particular, the term $\nabla a \times \vec E$ in the $rhs$ of electric field evolution equation Eq.~(\ref{eq:EOMvector2}) $\dot E_i = [...] + {e^2\over4\pi^2}(\nabla a \times \vec E)_i$, introduces a 'mixing' of the $E_i$ component, naturally living in the $i$-th direction, with the components $E_j, E_k$, naturally living along the transverse directions to the $i$th axis. As electric field components live naturally in between lattice sites (at the $links$), this will imply a mixture of orthogonal links. Some spatial averaging over neighboring position along the $i$-th axis will be needed, to force the term $\nabla a \times \vec E$ (in the $rhs$ of the equation) to live at the same location where $E_i$ lives (in the $lhs$ of the equation). This will create a non-local interaction, possibly preventing the development of an explicit iterative scheme to solve the resulting set of finite difference coupled equations. 

\end{enumerate}

%\subsection{Lattice axionic-coupling to order $\mathcal{O}(dx^2)$}
%\label{subsec:LatticeaxionCoupling}

In the following we will investigate various lattice versions of the interaction $\alpha\,\vec E\vec B$, determining their 'appropriateness' based on the ability of each lattice formulation to address the previous criteria $i)-iii)$. 

\subsection{Abelian gauge theory with a homogeneous axion}
\label{subsec:LatticeHomAxion} 

We will start considering in this section the simplest case of a gauge theory with a homogeneous axion $\alpha(x) = \alpha(t)$. As discussed in Sect.~\ref{sec:AbelianTheoryContinuum}, this problem can be mapped into the description of a gauge theory in the presence of a chemical potential $\mu = \dot \alpha$. Both in Sect.~\ref{subsubsec:Bianchi} and \ref{subsubsec:ExpScheme} we will stick to $a(x) = a(t)$, simply to make more transparent the discussion about the importance of achieving a good lattice representation of the Bianchi identities, as well of an explicit iterative scheme to solve the set of coupled lattice EOM. The conclusions that will be reached in Sects.~\ref{subsubsec:Bianchi}, \ref{subsubsec:ExpScheme} will be equally applicable to the case of a fully inhomogeneous axion $a(x) = a(t,{\bf x})$, which we will address in Sect.~\ref{subsec:LatticeInHomAxion}, building up from our previous findings on the homogeneous case.

\subsubsection{Lattice formulation of the Bianchi identities}
\label{subsubsec:Bianchi}

Let us first of all take the simplest possible approach, and attempt to describe $S_{ac}$ using the lattice definition of electric and magnetic fields introduced in Section~\ref{sec:LatticeBasics}, $E_i \equiv (\Delta_o^+ A_i - \Delta_i^+ A_o)$ and $B_i \equiv \epsilon_{ijk}\Delta_j^+A_k$, like
\begin{eqnarray}\label{eq:LatticeaxionCoupling1}
S_{ac}^{L(1)} \propto \sum_{\vec n,n_o} \alpha \sum_iE_iB_i = \sum_{\vec n,n_o} \alpha \sum_i (\Delta_o^+ A_i - \Delta_i^+ A_o)\epsilon_{ijk}\Delta_j^+A_k~.
\end{eqnarray} 
A priori, this looks like benign lattice operator, since it describes the continuum action up to order $\mathcal{O}(dx^2)$, as we showed already in Sect.~\ref{sec:LatticeBasics}: each term of $F_{\mn} \equiv (\Delta_\mu^+ A_\nu - \Delta_\nu^+ A_\nu)$ individually reproduces the continuum expression up to to order $\mathcal{O}(dx^2)$, when interpreting that $F_{\mn}$ lives at $n+{1\over 2}\hat\mu + {1\over 2}\hat\nu$. When varying this lattice action with respect $A_i$, we obtain (recall that we are assuming now $\alpha(x) = \alpha(t)$ homogeneous) a term $\propto \alpha[\Delta_o^-B_i-(\nabla^-\times\vec E)_i]$. When varying with respect to $A_o$, we obtain a term $\propto \alpha\sum\Delta_i^-B_i$. As expected, these terms resemble the continuum analogues $\propto \alpha(\partial_o{\vec B}-\vec\nabla\times \vec E)$ and $\propto \alpha\vec\nabla{\vec B}$ in Eqs.~(\ref{eq:EOMChPot2}), (\ref{eq:EOMChPot3}). However, contrary to the continuum analogues, they do not vanish. The reason is simple, the correct discrete version of $\vec\nabla{\vec B}$ is rather $\sum_i\Delta_i^+B_i = 0$, simply because the lattice magnetic field was defined in terms of forward derivatives, $B_i \equiv \epsilon_{ijk}\Delta_j^+A_k$. To obtain the desired result, one needs to take the divergence over $B_i$ with a forward derivative: as $\Delta_i^+$ and $\Delta_j^+$ commute, $\Delta_i^+\Delta_j^+A_k$ is symmetric in $i \leftrightarrow j$, and hence its contraction with $\epsilon_{ijk}$ vanishes, $\sum_i\Delta_i^+B_i = \epsilon_{ijk}\Delta_i^+\Delta_j^+A_k = 0$. However, the variation of Eq.~(\ref{eq:LatticeaxionCoupling1}) naturally led to $\sum\Delta_i^-B_i = \epsilon_{ijk}\Delta_i^-\Delta_j^+A_k \neq 0$, which does not vanish because $\Delta_i^-\Delta_j^+A_k$ is not symmetric in $i \leftrightarrow j$, given that $\Delta_i^-$ and $\Delta_j^+$ do not commute. Similarly, the term $[\Delta_o^-B_i-(\vec\nabla^-\times\vec E)_i \neq 0$] does not vanish. The appropriate version in the lattice of the Bianchi identity should rather be built as $[(\vec\Delta^+\times\vec E)_i - \Delta^+_o B_i] = 0$, as one can easily check that $\Delta_o^+B_i \equiv \epsilon_{ijk}\Delta_j^+\Delta_o^+A_k$ = $\epsilon_{ijk}\Delta_j^+(\Delta_o^+A_k-\Delta_k^+A_o)$ $\equiv (\vec\Delta^+\times\vec E)_i$. Variation of the lattice action Eq.~(\ref{eq:LatticeaxionCoupling1}) produced instead the expression $[(\vec\nabla^-\times\vec E)_i - \Delta_o^-B_i]$ = $\epsilon_{ijk}(\Delta_j^-\Delta_o^+ - \Delta_o^-\Delta_j^+) A_k \neq 0$, which does not vanish, simply because $\Delta_j^-\Delta_o^+$ and $\Delta_o^-\Delta_j^+$ do not commute.

As anticipated, generating the appropriated vanishing terms in the discrete EOM (due to the lattice version of the Bianchi identities), is not automatically granted. The problem arises because our choice of Eq.~(\ref{eq:LatticeaxionCoupling1}) as a lattice operator is actually not consistent. Note that even though it reproduces the continuum result to order $\mathcal{O}(dx^2)$, it consists however in the product of three fields, $\alpha, E_i$ and $B_i$ that live, not only in different lattice sites, also at different time steps. Whereas $\alpha$ lives at $(n_o,\vec n)$, $E_i$ lives at $(n_o+{1\over2},\vec n + {1\over2}\hat i)$ and $B_i$ lives at $(n_o,\vec n +{1\over2}\hat j + {1\over2}\hat k)$. To make consistent an action formed by the sum of several lattice operators, let us recall the rule we already discussed in Sec.~\ref{sec:LatticeBasics}, that all operators must reproduce the continuum limit to the same order, when expended around their natural site. It is therefore implicit in that statement, that each operator must have a well defined natural site where to live. This is precisely the reason why the previous operator Eq.~(\ref{eq:LatticeaxionCoupling1}) is inconsistent, as there is no natural site ascribed to it. The solution passes trough "symmetrizing" the factors in the operator, so that the factors built up from different fields, live nonetheless at the same lattice site. Let us define
\begin{eqnarray}
E_i^{(2)} \equiv {1\over2}(E_i+E_{i,-i})(l)\big|_{l \equiv n +{\hat 0\over2}} &\xrightarrow{~\vec x \equiv \vec n dx,~t \equiv (n_o+{1\over2})dt~} & E_i\left(\vec x, t\right) + \mathcal{O}(dx^2)\nn\\
E_i^{(4)} \equiv {1\over4}(E_i+E_{i,-i}+E_{i,-0}+E_{i,-i-0})(l)\big|_{l \equiv n} &\xrightarrow{~~~\vec x \equiv \vec n dx,~t \equiv n_o dt~~~~} & E_i(\vec x,t) + \mathcal{O}(dx^2)\nn\\
B_i^{(4)} \equiv {1\over4}(B_i+B_{i,-j}+B_{i,-k}+B_{i,-j-k})(l)\big|_{l \equiv n} &\xrightarrow{~~~\vec x \equiv \vec n dx,~t \equiv n_o dt~~~~} & B_i(\vec x,t) + \mathcal{O}(dx^2)\nn
\end{eqnarray}
and note that each of these fields can be expressed as
\begin{eqnarray}
E_i^{(2)} &\equiv& {1\over2}(2-dx\Delta_i^-)\Delta_o^+A_i\\
E_i^{(4)} &\equiv& {1\over4}(\Delta_o^+ + \Delta_o^-)(2-dx\Delta_i^-)A_i\\
B_i^{(4)} &\equiv& {1\over4}\sum_{j,k}\epsilon_{ijk}(\Delta_j^+ + \Delta_j^-)(2-dx\Delta_k^-)A_k
\end{eqnarray}
%where we have defined $\delta_i^{\pm} \equiv dx\Delta_i^\pm$. 
For convenience we also define
\begin{eqnarray}
%E_i^{(8)} &\equiv& {1\over2}\left(E_i^{(4)}(l)\big|_{l \equiv n}+E_{i,+i}^{(4)}(l)\big|_{l \equiv n+\hat i}\right)\,,\\
%B_i^{(8)} &\equiv& {1\over2}\left(B_i^{(4)}(l)\big|_{l \equiv n}+B_{i,+i}^{(4)}(l)\big|_{l \equiv n+\hat i}\right)
E_i^{(8)} &\equiv& {1\over2}\left(E_i^{(4)}+E_{i,+i}^{(4)}\right) = {1\over8}(2+dx\Delta_i^+)(2-dx\Delta_i^-)(\Delta_o^+ + \Delta_o^-)A_i\,,\\
B_i^{(8)} &\equiv& {1\over2}\left(B_i^{(4)}+B_{i,+i}^{(4)}\right) = {1\over8}\sum_{j,k}\epsilon_{ijk}(2+dx\Delta_i^+)(\Delta_j^+ + \Delta_j^-)(2-dx\Delta_k^-)A_k\,,
\end{eqnarray}
which reproduce the continuum as
\begin{eqnarray}
E_i^{(8)} &\xrightarrow{~\vec x \equiv (\vec n + {1\over2}\hat i) dx,~t \equiv n_o dt~} & E_i(\vec x,t) + \mathcal{O}(dx^2)\\
B_i^{(8)} &\xrightarrow{~\vec x \equiv (\vec n + {1\over2}\hat i) dx,~t \equiv n_o dt~} & B_i(\vec x,t) + \mathcal{O}(dx^2)
\end{eqnarray}

A 'symmetrized' operator that reproduces the continuum expression of $S_{ac}$ at $l = (\vec n, n_o + {1\over2})$ to order $\mathcal{O}(dx^2)$, can be easily proposed based on the above expressions,
\begin{eqnarray}\label{eq:LatticeaxionCoupling2}
S_{ac}^{L(2)} &\propto& \sum_{\vec n,n_o} \alpha \sum_iE_i^{(4)}B_i^{(4)}\\
&=& \sum_{\vec n,n_o} {\alpha\over16} \sum_{i,j,k} \epsilon_{ijk}[(\Delta_o^+ + \Delta_o^-)(2-dx\Delta_i^-)A_i][(\Delta_j^+ + \Delta_j^-)(2-dx\Delta_k^-)A_k]\,.\nonumber
\end{eqnarray} 
Varying Eq.~(\ref{eq:LatticeaxionCoupling2}) with respect $A_i$, produces a term in the discrete EOM of the gauge field as $\alpha\left(\sum_{j,k}\epsilon_{ijk}(\Delta^+_j + \Delta^-_j)E^{(8)}_k\right.$  $- \left.(\Delta_o^+ +\Delta_o^-)B_i^{(8)}\right)$, which resembles the continuum term $\alpha(\epsilon_{ijk}\partial_jE_k-\dot B_i)$ in Eq.~(\ref{eq:EOMChPot2}). Whereas the latter vanishes thanks to the Bianchi identity in the continuum $\vec\nabla\times \vec E = \dot {\vec B}$, it can be shown, with a bit of algebra, that a lattice version of this identity holds as $\sum_{j,k}\epsilon_{ijk}(\Delta^+_j + \Delta^-_j)E^{(8)}_k$ = $(\Delta_o^+ +\Delta_o^-)B_i^{(8)}$. This implies that the new term encountered in the gauge field discrete EOM just vanishes. Similarly, when varying Eq.~(\ref{eq:LatticeaxionCoupling2}) with respect $A_o$, we produce a new term $\alpha\sum_i(2+dt\Delta^+_t)\Delta_i^-B_i^{(8)}$, which again resembles the term in the continuum $\alpha\partial_iB_i$ in Eq.~(\ref{eq:EOMChPot3}), which vanishes due to the Bianchi identity $\vec\nabla \vec B = 0$. With a bit of algebra, it can be shown that the analogous lattice identity reads $\Delta_i^-B_i^{(8)} = (\Delta_i^++\Delta_i^-)B_i^{(4)} = 0$, so that the new term encountered in the analogous discrete Gauss law, just vanishes. This completes the proof that the new operator Eq.~(\ref{eq:LatticeaxionCoupling2}) represents a good lattice candidate from which to derive [together with action Eq.~(\ref{eq:ActionLAH})] a set of coupled finite different equations reproducing correctly the functional form of continuum EOM. As we anticipated, there is however another problem yet to be circumvented, related to the solubility of a set of coupled finite different equations.

\subsubsection{Explicit scheme for real time evolution}
\label{subsubsec:ExpScheme}

The operator Eq.~(\ref{eq:LatticeaxionCoupling2}) proposed to represent an axionic coupling, exhibits various features: $i)$ it reproduces correctly the continuum term to order $\mathcal{O}(dx^2)$, and $ii)$ it reproduces correctly a lattice version of the Bianchi identities, so that the discrete EOM reproduce correctly the functional form of the dynamical Eqs.~(\ref{eq:EOMChPot1})-(\ref{eq:EOMChPot4}) in the continuum. In fact, varying Eq.~(\ref{eq:LatticeaxionCoupling2}) with respect to the gauge fields, generates a term $\propto {1\over2}\left(\mu_{-0}B_{i}^{(8)}+\mu B_{i,+0}^{(8)}\right)$, which reduces correctly to the continuum term $\mu\vec B$ in Eq.~(\ref{eq:EOMChPot2}), to order $\mathcal{O}(dx^2)$. At the same time, varying Eq.~(\ref{eq:LatticeaxionCoupling2}) with respect to $\alpha(t)$, generates a term $\propto {1\over2}(E_{i}^{(2)}+E_{i,-0}^{(2)})B_i^{(4)}$ sourcing the chemical potential, which again reduces correctly to the continuum source term $\propto \vec E \vec B$ in Eq.~(\ref{eq:EOMChPot4}), to order $\mathcal{O}(dx^2)$. The set of coupled discrete equations one obtains, cannot be put however in an explicit iterative scheme, because in order to find $E_i$ we need $\mu$ and $A_{i,+0}$ (to obtain $B_{i,+0}^{(8)}$), and at the same time to find $A_{i,+0}$ and $\mu$ we need $E_i$. One would need to express the term $\mu B_{i,+0}^{(8)}$ in the gauge field EOM in terms of $E_i$, and then solve for $E_i$, but this would complicate the equations unnecessary\footnote{Actually, it is not even feasible, in principle, to solve this system, as e.g.~$B_{1,+0}^{(8)}$ depends $A_{2,+\hat 1 + \hat 0}$ and $A_{3,+\hat 1 + \hat 0}$, which depend respectively on $E_{2,+1}$ and $E_{3,+1}$. So at the end, the equation for the $\hat 1$-component of the electric field depends on $E_1$ but also on $E_{2,+1}$ and $E_{3,+1}$, and analogously for the other electric field component equations. Therefore, one cannot just simply solve for the $E_i$ components at a given time step and lattice site, as we would need to know the electric field at all other lattice sites. We will actually also encounter this problem in Sec.~\ref{subsec:LatticeInHomAxion} when solving the system for an inhomogenous axion-like field, so we postpone any further discussion on this till then.}. A simpler solution consists in choosing another lattice operator that upon variation over the gauge fields, prevents the duplication of the $\sim\mu B_i$ terms at the two different times. The duplication of these terms originated from Eq.~(\ref{eq:LatticeaxionCoupling2}) due to the fact that the electric field $E_i^{(4)}$ multiplying $\alpha B_i^{(4)}$ in the operator, is equivalent to the sum of electric fields at two time steps, $E_i^{(4)} \equiv {1\over 2}(E_i^{(2)}+E_{i,-0}^{(2)})$. Hence, when varying with respect to $A_i$, two terms of the form $\sim \mu B_i$ are generated, $\left(\mu_{-0}B_{i}^{(8)}+\mu B_{i,+0}^{(8)}\right)$, each evaluated at a different time.

It is easy to build some lattice operator that verifies the last requisite, i.e.~not involving the sum of two gauge field conjugate momenta (i.e.~electric fields) at different times, like for example
\begin{eqnarray}\label{eq:LatticeaxionCoupling3}
S_{ac}^{L(3)} &\propto& \sum_{\vec n,n_o} \alpha \sum_iE_i^{(2)}B_i^{(4)} \\
&=& \sum_{\vec n,n_o} {\alpha\over8} \sum_{i,j,k} \epsilon_{ijk}[(2-dx\Delta_i^-)(\Delta_o^+ A_i)][(\Delta_j^+ + \Delta_j^-)(2-dx\Delta_k^-)A_k]\,.\nonumber
\end{eqnarray} 
This operator fails however in 'symmetrizing' the expression for $\vec E \cdot \vec B$ around a common time: as $B_i^{(4)}$ lives at integer times $n_o$ whereas $E_i^{(2)}$ lives at time semi-integer timEs $(n_o+{1\over2})$, the lattice equivalent to the terms that should be vanishing due to the discretized Bianchi identities, will fail to vanish, as it happened already with the operator $S_{ac}^{L(1)}$ Eq.~(\ref{eq:LatticeaxionCoupling1}). 

A better solution is found if the lattice operator $\sim \alpha(E\cdot B)$ is built such that each element $\alpha$, $\vec E$ and $\vec B$, live separately in a given common space-time site. In other words, contrary to the operator $S_{ac}^{L(3)}$ Eq.~(\ref{eq:LatticeaxionCoupling3}), where the electric and magnetic fields lived at different time steps separated away by half step ${dt\over2}$, now both $\vec E$ and $\vec B$ must live at the same time step. As the natural time where electric fields live is $(n_o + {1\over2})dt$, and given that we do not want to sum over electric fields at different times, the natural option will be to make the magnetic field to live in the same semi-integer time. There are 3 options for this,
\begin{eqnarray}\label{eq:LatticeaxionCoupling4}
S_{ac}^{L(4)} &\propto& \sum_{\vec n,n_o} \alpha \sum_iE_i^{(2)}(B_i^{(4)}+B_{i+0}^{(4)}) \equiv \sum_{\vec n,n_o} \sum_i(\alpha E_i^{(2)}+\alpha_{-0}E_{i,-0}^{(2)})B_i^{(4)}\nn
&\propto& \sum_{\vec n,n_o} {\alpha\over16} \sum_{i,j,k} \epsilon_{ijk}[(2-dx\Delta_i^-)(\Delta_o^+ A_i)][(\Delta_j^+ + \Delta_j^-)(2-dx\Delta_k^-)(2+dt\Delta_o^+)A_k]\,,\\
\label{eq:LatticeaxionCoupling5}
S_{ac}^{L(5)} &\propto& \sum_{\vec n,n_o} \alpha_{+0} \sum_iE_i^{(2)}(B_i^{(4)}+B_{i+0}^{(4)}) \equiv \sum_{\vec n,n_o} \sum_i(\alpha_{+0} E_i^{(2)}+\alpha E_{i,-0}^{(2)})B_i^{(4)}\nn
&\propto& \sum_{\vec n,n_o} {\alpha_{+0}\over16} \sum_{i,j,k} \epsilon_{ijk}[(2-dx\Delta_i^-)(\Delta_o^+ A_i)][(\Delta_j^+ + \Delta_j^-)(2-dx\Delta_k^-)(2+dt\Delta_o^+)A_k]\,,\\
\label{eq:LatticeaxionCoupling6}
S_{ac}^{L(6)} &\propto& \sum_{\vec n,n_o} (\alpha+\alpha_{+0}) \sum_iE_i^{(2)}(B_i^{(4)}+B_{i+0}^{(4)}) \equiv \sum_{\vec n,n_o} \sum_i[(\alpha+\alpha_{+0})E_i^{(2)}+(\alpha+\alpha_{-0})E_{i,-0}^{(2)}]B_i^{(4)}\nn
&\propto& \sum_{\vec n,n_o} {(\alpha+\alpha_{+0})\over32} \sum_{i,j,k} \epsilon_{ijk}[(2-dx\Delta_i^-)(\Delta_o^+ A_i)][(\Delta_j^+ + \Delta_j^-)(2-dx\Delta_k^-)(2+dt\Delta_o^+)A_k]\,,\nn
\end{eqnarray}
where $\equiv$ should be understood as an equivalence modulo an additive constant shift. Varying each of these action terms with respect to the gauge fields, we obtain correct vanishing versions of $\vec\nabla\vec B$ and $(\dot{\vec B} - \vec\nabla \times \vec E)$ in the discrete equations of motion due to the lattice Bianchi identities, and generate the following terms [recall that here we still consider $\alpha(x) = \alpha (t)$],
\begin{eqnarray}
\begin{array}{cclcc}
{\rm Action} & & A_i~{\rm EOM~term} & & \mu~{\rm source~term}\vspace*{0.25cm}\\
\delta S_{ac}^{L(4)}  = 0 & ~~~\Rightarrow~~~ & -{e^2\over 4\pi^2}\mu_{-0} B_i^{(8)} & \,, ~~& E_i^{(2)}(B_i^{(4)}+B_{i,+0}^{(4)})\label{eq:VariationSL4}\\
\delta S_{ac}^{L(5)}  = 0 & ~~~\Rightarrow~~~ & -{e^2\over 4\pi^2}\mu B_i^{(8)} & \,, ~~& E_{i,-0}^{(2)}(B_i^{(4)}+B_{i,-0}^{(4)})\label{eq:VariationSL5}\\
\delta S_{ac}^{L(6)}  = 0 & ~~~\Rightarrow~~~ & -{e^2\over 4\pi^2}{1\over 2}(\mu + \mu_{-0})B_i^{(8)} &\,, ~~& {1\over2}[E_i^{(2)}(B_i^{(4)}+B_{i,+0}^{(4)})+E_{i,-0}^{(2)}(B_i^{(4)}+B_{i,-0}^{(4)})]\label{eq:VariationSL6}
\end{array}
\end{eqnarray} 
From here we see that $S_{ac}^{L(6)}$ Eq.~(\ref{eq:LatticeaxionCoupling6}) does not allow for an explicit scheme of iteration\footnote{In this occasion this occurs because even though there is a single electric field $E_{i}^{(2)}$ [hence defined at its natural time $(n_o+1/2)dt$], the latter is multiplied by $(\alpha+\alpha_{+0})$ with the axion-like field living at two different times.}, similarly as what it happened with $S_{ac}^{L(2)}$ Eq.~(\ref{eq:LatticeaxionCoupling2}). Only actions $S_{ac}^{L(4)}$ Eq.~(\ref{eq:LatticeaxionCoupling4}) and $S_{ac}^{L(5)}$ Eq.~(\ref{eq:LatticeaxionCoupling5}) allow allow for an explicit scheme of iteration, and in fact produce essentially an equivalent set of coupled discrete EOM. 

In conclusion, although all operators $S_{ac}^{L(1)}$-$S_{ac}^{L(6)}$ reproduce correctly the continuum gauge-axionic interaction to order $\mathcal{O}(dx^2)$, $S_{ac}^{L(1)}$ Eq.~(\ref{eq:LatticeaxionCoupling1}) and $S_{ac}^{L(3)}$ Eq.~(\ref{eq:LatticeaxionCoupling3}) fail to generate vanishing terms in the discrete EOM from the lattice Bianchi identities, whereas $S_{ac}^{L(2)}$ Eq.~(\ref{eq:LatticeaxionCoupling2}) and $S_{ac}^{L(6)}$ Eq.~(\ref{eq:LatticeaxionCoupling6}) produce correctly null terms due to the lattice Bianchi identities, but fail to generate explicit iterative schemes for solving the set of coupled finite difference EOM. Only $S_{ac}^{L(4)}$ Eq.~(\ref{eq:LatticeaxionCoupling4}) and $S_{ac}^{L(5)}$ Eq.~(\ref{eq:LatticeaxionCoupling5}), produce correctly vanishing terms due to the lattice Bianchi identities, while maintaining an explicit iterative scheme for solving the set of lattice EOM. Actually, the functional form of $S_{ac}^{L(4)}$ Eq.~(\ref{eq:LatticeaxionCoupling4}) [equivalently $S_{ac}^{L(5)}$ Eq.~(\ref{eq:LatticeaxionCoupling5})] indicates us something important: the natural time steps where an shift-symmetric field lives are also semi-integer times, and not integer steps as for ordinary scalar fields. Pseudo-scalar fields (independently of whether they are homogeneous as we have just required so far) must live in semi-integer times in the lattice. In light of this, we can interpret $S_{ac}^{L(4)}$ Eq.~(\ref{eq:LatticeaxionCoupling4}) and $S_{ac}^{L(5)}$ Eq.~(\ref{eq:LatticeaxionCoupling5}) as equivalent descriptions of a final suitable action. Let us notice as well that, as a consequence of this, the kinetic term of the axion should be also defined differently compared to ordinary scalar fields, since we want to obtain a finite difference evaluated at integer times. 

Putting all together, the discretized action from where  the dynamical effects of the presence of a homogeneous axion to be derived reads
\begin{eqnarray}\label{eq:LatticeHomAxionAction}
S_{{\alpha(t)}}^{L} &=& \Delta t \Delta x^3\sum_{\vec n,n_o}\left\lbrace {M^2\over 2c_s^2}\left(\Delta_o^-\alpha\right)^2 + {1\over 4\pi^2}\alpha\sum_i {1\over2}E_i^{(2)}\left(B_i^{(4)}+B_{i,+0}^{(4)}\right)\right\rbrace \nn
&\equiv& \Delta t \Delta x^3\sum_{n_o}\left\lbrace M^2{N^3\over 2c_s^2}\left(\Delta_o^-\alpha\right)^2 + {1\over 4\pi^2}\alpha\sum_{\vec n} {1\over2}\vec E^{(2)}\left(\vec B^{(4)} + \vec B_{+0}^{(4)}\right)\right\rbrace\,,
%\\&=& {e^2\over 4\pi^2}\sum_{\vec n,n_o} {\alpha(l)\over16} \sum_{i,j,k} \epsilon_{ijk}[(2-\delta_i^-)(\Delta_o^+ A_i)](l)[(\Delta_j^+ + \Delta_j^-)(2-\delta_k^-)(2+\delta_o^+)A_k](l)\Big|_{l \equiv (n,n_o+\hat 0/2)}\,,
\end{eqnarray}
where it is important to note that both $\alpha$ and $A_o$ live at space-time sites $(n_o + {1\over2},\vec n)$, whereas $A_i$ live at $(n_o,\vec n + {\hat i\over2})$. Consequently, $\mu \equiv \Delta_o^-\alpha$ lives naturally at $(n_o,n)$, $E_i$ at $(n_o+{1\over2},\vec n+{1\over2}\hat i)$, $E_i^{(2)}$ at $(n_o+{1\over2},\vec n)$, whereas $B_i$ lives at $(n_o,n+{1\over2}\hat j+{1\over2}\hat k)$, and $B_i^{(4)}$ at $(n_o,\vec n)$. 

It is perhaps relevant to stress that, at the end, the need for all the factors multiplying within a given operators living at the same space-time site $(l_o,\vec l)$, is a crucial aspect for determining the right functional form the lattice operator. As we saw, however, this is not enough, as $S_{ac}^{L(2)}$ Eq.~(\ref{eq:LatticeaxionCoupling2}) verifies that, with $(l_o,\vec l) = (n_o,\vec n)$, but still has problems. One must first recognize the natural space-time site where the operator lives. As we do not want to have the sum of electric fields at different times in the operator, the appropriate representation for the electric field is $E_{i}^{(2)}$, which lives in integer lattice sites, but semi-integer times, i.e.~$(l_o,\vec l) = (n_o+{1\over2},\vec n)$. This implies, correspondingly, that the appropriate magnetic field representation must be ${1\over2}(B_{i}^{(4)}+B_{i,+0}^{(4)})$, so that it also lives in $(l_o,\vec l) = (n_o+{1\over2},\vec n)$. The linear part in the (pseudo-)scalar field $\alpha$, can then be made meaningfully defined only if $\alpha$ also lives at $(l_o,\vec l) = (n_o+{1\over2},\vec n)$.

The final set of equations of motion obtained from varying $S_{\rm AH}^{L} + S_{{\alpha(t)}}^{L}$ [Eq.~(\ref{eq:ActionLAH}) + Eq.~(\ref{eq:LatticeHomAxionAction})], mimicking a system with chemical potential $\mu \equiv \Delta_o^-\alpha$ at finite temperature $T$, where recall that $M = T$ and $c_s^2 = 12$, are (we use the Coulomb gauge $A_o = 0$, so that $U_o = 1$)
\begin{eqnarray}
\vspace*{1.5cm}
{\rm Equation} \hspace*{7.8cm} {\rm Natural~Site}\hspace*{1.3cm}\nonumber\vspace*{0.5cm}\\
\begin{array}{rclcl}
\pi &\equiv& \Delta_o^+\varphi\,, & \rightarrow & l \equiv (n_o+{1\over2},\vec n)\\
E_i &\equiv& \Delta_o^+A_i\,, & \rightarrow & l \equiv (n_o+{1\over2},\vec n+{1\over2}\hat i)\\
\mu &\equiv& \Delta_o^-\alpha\,, & \rightarrow & l \equiv (n_o,\vec n)\\
\Delta_o^-\pi &=& \sum_iD_i^-D_i^+\varphi - V_{,\varphi^*} = 0\, & \rightarrow & l \equiv (n_o,\vec n)\\
\Delta_o^- E_i &=& {2e^2}{\rm Im}\lbrace\varphi^*D_i^+\varphi\rbrace -\sum_{j,k}\epsilon_{ijk}\Delta_{j}^-B_k%\sum_{j}\left(\Delta_j^-\Delta_j^+A_i - \Delta_i^+ \Delta_j^-A_j\right) 
 - {e^2\over 4\pi^2}\mu B_i^{(8)}\,, & \rightarrow & l \equiv (n_o,\vec n + {1\over2}\hat i)\\
\sum_i \Delta_i^- E_i &=& {2e^2}{\rm Im}\lbrace\varphi^*\pi\rbrace\hV(\rm Gauss\,\,Constraint)\,, & \rightarrow & l \equiv (n_o+{1\over2},\vec n)\\
\Delta_o^+\mu &=& {3\over \pi^2} {1\over T^2}{1\over N^3}\sum_{\vec n} {1\over2}\sum_i E_i^{(2)}(B_i^{(4)}+B_{i,+0}^{(4)})\,, & \rightarrow & l \equiv (n_o+{1\over2},\vec n)
\end{array}\label{eq:EOMlatticeChemPot}
\end{eqnarray} 
Let us emphasize that, only thanks to the fact that $\alpha$ lives at $(l_o,l) = (n_o+{1\over2},n)$, so that $\mu$ lives at $(l_o,\vec l) = (n_o,\vec n)$, we can make real sense of these discrete equations. Only thanks to this interpretation, the terms within each of the above equations live at a given common natural space-time site (specifically indicated above in the $rhs$ of each equation), around which we can expand each equation and reproduce the continuum analogue Eqs.~(\ref{eq:EOMChPot1})-(\ref{eq:EOMChPot4}), up to order $\mathcal{O}(dx^2)$. 

\subsection{Abelian gauge theory with an inhomogeneous axion}
\label{subsec:LatticeInHomAxion}

Let us just recall action Eq.~(\ref{eq:ActionContinuum}) for a fully inhomegeneous axion-like field $a(x)$,
\begin{eqnarray}\label{eq:ActionContinuumIII}
S = \int d^4x\left(-\mathcal{L}_{\varphi} + {1\over 2e^2}\left({\vec E}^2- {\vec B}^2\right) + {1\over2 c_s^2}{\dot a}^2 - {1\over2}|\nabla a|^2 + {1\over 4\pi^2}{a\over M}\vec{E}\vec{B}\right)\,,
\end{eqnarray}
with $\mathcal{L}_{\varphi} \equiv (D_0\varphi)^*(D_0\varphi) - (\vec{D}\varphi)^*(\vec{D}\varphi) + V(\varphi^*\varphi)$ characterizing the Higgs sector. Variation of the action produces the EOM (imposing already the Bianchi identities)
\begin{eqnarray}\label{eq:EOMaxion1}
D_oD_o\varphi &=& \vec D \vec D\varphi - V_{,|\varphi|^2}\varphi\,,\\
\label{eq:EOMaxion2}
\dot{\vec E} + \vec\nabla \times \vec B %+ {e^2\over 4\pi^2}{a\over M}(\dot{\vec B} - \vec\nabla \times \vec E) 
&=& e^2\vec{J} - {e^2\over 4\pi^2 M}{\dot a}\vec B + {e^2\over 4\pi^2 M}{\vec\nabla a}\times \vec E\,,\\
\label{eq:EOMaxion3}
\vec\nabla \vec E %{e^2\over 4\pi^2}{a\over M}\vec\nabla\vec B 
&=& e^2\rho - {e^2\over 4\pi^2 M}{\vec\nabla a}\cdot \vec B\,,\\
\label{eq:EOMaxion4}
\ddot a &=& c_s^2{\vec\nabla}^2a + {c_s^2\over 4\pi^2 M}\vec E \cdot \vec B\,,
\end{eqnarray}
Various differences arise with respect the set of Eqs.~(\ref{eq:EOMChPot1})-(\ref{eq:EOMChPot4}) describing an homogeneous field $a(x) = a(t)$. First, the gauge field EOM Eq.~(\ref{eq:EOMaxion2}) includes now a term $\propto {\vec\nabla a}\times \vec E$. Secondly, the Gauss law Eq.~(\ref{eq:EOMaxion3}) includes a term $\propto {\vec\nabla a}\cdot \vec B$. Third, the axion dynamics follows a (sourced) wave equation, where $c_s^2$ now plays the role of a real propagation speed. In other words, there are new terms affecting the dynamics (except in the higgs sector) due to the spatial dependence of $a(x)$. Action Eq.~(\ref{eq:ActionContinuumIII}) and the corresponding set of Eqs.~(\ref{eq:EOMaxion1})-(\ref{eq:EOMaxion4}), cannot be used anymore to describe the problem of an Abelian gauge theory with a chemical potential. They rather describe an Abelian gauge theory in the presence to a fully inhomogeneous axion field. We shall understand that from now on we are dealing with such scenario. 

\subsubsection{Implicit scheme for real time evolution}
\label{subsubsec:ImpScheme}

In order to find a lattice formulation for the interaction $a(x)\tilde F_\mn F^\mn \propto a\vec E\cdot \vec B$ where $a(x)$ is now an inhomogeneous field, we can proceed in the same manner as in Sect.~\ref{subsec:LatticeHomAxion}, when $a(x) = a(t)$ was simply considered a homogeneous field. The same considerations apply now in order to find an appropriate lattice representation of the gauge-axion interaction. We can thus survey the same lattice implementations $S_{ac}^{L(1)}$-$S_{ac}^{L(6)}$ proposed in Sec.~\ref{subsec:LatticeHomAxion}, as we know they all reproduce correctly the continuum interaction $a(x)\tilde F_\mn F^\mn$ to order $\mathcal{O}(dx^2)$. The spatial dependence of $a(x)$ does not change the fact that $S_{ac}^{L(1)}$ Eq.~(\ref{eq:LatticeaxionCoupling1}) and $S_{ac}^{L(3)}$ Eq.~(\ref{eq:LatticeaxionCoupling3}), fail again to generate vanishing terms in the discrete EOM due to the lattice Bianchi identities. We also find that $S_{ac}^{L(2)}$ Eq.~(\ref{eq:LatticeaxionCoupling2}) and $S_{ac}^{L(6)}$ Eq.~(\ref{eq:LatticeaxionCoupling6}) produce correctly null terms due to the lattice Bianchi identities, but fail again to generate explicit iterative schemes for solving the set of coupled finite different EOM. The expressions $S_{ac}^{L(4)}$ Eq.~(\ref{eq:LatticeaxionCoupling4}) and $S_{ac}^{L(5)}$ Eq.~(\ref{eq:LatticeaxionCoupling5}) produce correctly vanishing terms in the gauge field discrete EOM due to the lattice Bianchi identities. However, contrary to the homogeous $a(x) = a(t)$ case, $S_{ac}^{L(4)}$-$S_{ac}^{L(5)}$ do not lead now to an explicit iterative scheme for solving the set of lattice EOM when $a(x) = a(t,{\bf x})$ is inhomogeous.

Let us see this more in detail, considering $S_{ac}^{L(4)}$ Eq.~(\ref{eq:LatticeaxionCoupling4}) as the lattice representation of the axion-gauge field interaction [$S_{ac}^{L(5)}$ Eq.~(\ref{eq:LatticeaxionCoupling5}) is essentially equivalent]. We write the final lattice action mimicking to order $O(dx^2)$ the continuum action Eq.~(\ref{eq:ActionContinuumIII}) as
\begin{eqnarray}\label{eq:LatticeaxionActionFinal}
S = S_{AH} + S_{axion} &=& \Delta t \Delta x^3\sum_{\vec n,t} \lb %\frac{1}{2}\Delta_o^+\phi - \frac{1}{2}\sum_j\Delta_j^+\phi + 
(D_o^{+}\varphi)^\dag(D_o^{+}\varphi) - \sum_j(D_j^{+}\varphi)^\dag(D_j^{+}\varphi) - V(\varphi\varphi^*,\phi) \right. \nn 
&& ~~~~+~ \frac{1}{2e^2}\sum_{j}\left(\Delta_o^+A_i-\Delta_i^+A_o\right)^2  - \frac{1}{4e^2}\sum_{i,j}(\Delta_i^+A_j-\Delta_j^+A_i)^2\\
&& +~\left. {1\over 2c_s^2}\left(\Delta_o^-a\right)^2 - {1\over 2}\sum_i(\Delta_i^+a)^2 + {1\over 4\pi^2}{a\over M}\sum_i {1\over2}E_i^{(2)}\left(B_i^{(4)}+B_{i,+0}^{(4)}\right)\right\rbrace\nonumber
\end{eqnarray}
Varying this action, we obtain a set of finite difference coupled equations,
\begin{eqnarray}
%\vspace*{1.5cm}
%{\rm Equation} \hspace*{7.8cm} {\rm Natural~Site}\hspace*{1.3cm}\nonumber\vspace*{0.5cm}\\
\begin{array}{rcl}
\pi_\varphi &\equiv& \Delta_o^+\varphi\,,\\
E_i &\equiv& \Delta_o^+A_i\,,\\
\pi_a &\equiv& \Delta_o^-a\,,\\
\Delta_o^-\pi_\varphi &=& \sum_iD_i^-D_i^+\varphi - V_{,\varphi^*}\,,\\
\Delta_o^- E_i &=& {2e^2}{\rm Im}\lbrace\varphi^*D_i^+\varphi\rbrace -\sum\limits_{j,k}\epsilon_{ijk}\Delta_{j}^-B_k - {e^2\over 4\pi^2 M}{1\over2}(\mu B_i^{(4)} + \mu_{+i} B_{i,+i}^{(4)}) \\
%\sum_{j}\left(\Delta_j^-\Delta_j^+A_i - \Delta_i^+ \Delta_j^-A_j\right) 
 && +~ {e^2\over 4\pi^2 M}{1\over8}(2+dx\Delta_i^+)\sum\limits_{\pm}\sum\limits_{jk}\epsilon_{ijk}\lb[(\Delta_j^\pm a) E_{k,\pm j}^{(2)}]+[(\Delta_j^\pm a) E_{k,\pm j}^{(2)}]_{_{-0}}\rb \,,\\
\sum\limits_i \Delta_i^- E_i &=& {2e^2}{\rm Im}\lbrace\varphi^*\pi\rbrace - {e^2\over 4\pi^2 M}{1\over8}\sum\limits_{\pm}\sum\limits_i (\Delta_i^{\pm}a)(B_i^{(4)}+B_{i,+0}^{(4)})_{\pm i}\,,\\
\Delta_o^+\pi_a &=& c_s^2\sum\limits_i\Delta_i^-\Delta_i^+a + {3\over \pi^2} {c_s^2\over M}{1\over N^3}\sum\limits_{\vec n} \sum\limits_i {1\over2}E_i^{(2)}(B_i^{(4)}+B_{i,+0}^{(4)})\,,
\end{array}\label{eq:EOMlatticeAxion}
\end{eqnarray} 
which reproduce to order $\mathcal{O}(dx^2)$ the set of continuum Eqs.~(\ref{eq:EOMaxion1})-(\ref{eq:EOMaxion4}), when expanding each equation around its natural lattice site. Note that  the natural sites ascribed to each discrete EOM in Eqs.~(\ref{eq:EOMlatticeAxion}) coincide with those listed in the $rhs$ of Eqs.~(\ref{eq:EOMlatticeChemPot}), so we do not repeat them here. A simple inspection of the $lhs$ of each equation suffices anyways to determine these sites, e.g.~knowing that the lattice representation of the electric field $E_i = \Delta_o^+A_i$ lives at $l = (n_o + {1\over2},\vec n + {1\over2}\hat i)$, then equation $E_i \equiv [...]$ must be expanded around $x = ((n_o + {1\over2})dt,(\vec n + {1\over2}\hat i)dx)$, $\Delta_o^-E_i = [...]$ around $x = (n_odt,(\vec n + {1\over2}\hat i)dx)$, etc.  

Unfortunately, the set of finite difference coupled Eqs.~(\ref{eq:EOMlatticeAxion}) cannot be solved with an explicit scheme. In fact, these equations can only be formally solved by a non-local solution, even though we started from a local Lagrangian~Eq.~(\ref{eq:LatticeaxionActionFinal}). The reason for this is the following. Say we consider the $\hat 1$-component of the curl product involving the electric field in the $rhs$ of the gauge field EOM $\Delta_o^-E_{1} = [...]$, i.e.~$\sim [(\Delta a) \times \vec E^{(2)}]_1$. This term depends on $E_{2,+1}$ and $E_{3,+1}$, so the equation to update the $\hat 1$-component of the electric field depends on $E_{2,+1}$ and $E_{3,+1}$, and analogously for the other electric field component equations. In other words, there is a first-neighbour coupling of the electric field components. This simply makes impossible to solve explicitly for the $E_i$ components at a given lattice site (in a given time step), as we would need to know the electric field components at all other lattice sites (at the new time step). %Due to this, it does not even seem feasible to solve iteratively this system of equations, not even by some implicit method. 

The origin of this problem lies of course in the form of the interaction $\sim a\vec E \vec B$. In the moment $a(x) = a(\vec x,t)$ is inhomogeneous, this brings up the term $\vec \nabla a \times \vec E$ in the $rhs$ of the gauge field continuum EOM Eq.~(\ref{eq:EOMaxion2}), $\dot{\vec E} = [...]$. As in the lattice $E_i$ lives at $\vec n + {1\over 2}\hat i$, the equivalent lattice expression representing the continuum term $\sim (\vec \nabla a \times \vec E)$ must be evaluated at both $\vec n$ and $\vec n + \hat i$. This explains in fact the $(2+dx\Delta^+_i)$ operation in $rhs$ of the equivalent discrete EOM $\Delta_o^-E_i = [...]$ within Eqs.~(\ref{eq:EOMlatticeAxion}). This turns non-local the set of coupled equations in finite differences Eqs.~(\ref{eq:EOMlatticeAxion}), hence making it unfeasible to solve them iteratively by an explicit scheme at each lattice site. 

In conclusion, it does not seem possible to find a set of discrete equations reproducing to order $\mathcal{O}(dx^2)$ the dynamics of an Abelian gauge theory in the presence of a general axion-like field $a(x) = a(t,{\bf x})$, and at the same time being solvable by an explicit scheme. An approximate way around this difficulty can however be obtained by an implicit scheme as follows. Let us write the discrete equation evolving the electric fields, but writing down only the terms that prevented us from achieving an explicit scheme
\begin{eqnarray}\label{eq:ElecFldEOM}
E_{i,+{\hat 0\over 2}} &=& E_{i,-{\hat 0\over 2}} + \Delta t\,[...] \\
&& +~ \Delta t\,\left({e^2\over 4\pi^2 M}{1\over8}(2+dx\Delta_i^+)\sum\limits_{\pm}\sum\limits_{jk}\epsilon_{ijk}\lb[(\Delta_j^\pm a) E_{k,\pm j}^{(2)}]_{+{\hat 0\over 2}}+[(\Delta_j^\pm a) E_{k,\pm j}^{(2)}]_{-{\hat 0\over 2}}\rb\right),\nonumber
\end{eqnarray}
where $[...]$ represents all the terms in the $rhs$ of the discrete equation which involve only the amplitude (or spatial gradients) of the gauge field $A_i$, but not the electric field. Now let us suppose that we approximate the electric field term in the $rhs$ of Eq.~(\ref{eq:ElecFldEOM}) as
\begin{eqnarray}\label{eq:ElecFldApprox}
\lb[(\Delta_j^\pm a) E_{k,\pm j}^{(2)}]_{+{\hat 0\over 2}}+[(\Delta_j^\pm a) E_{k,\pm j}^{(2)}]_{-{\hat 0\over 2}}\rb ~~ \simeq ~~ 2[(\Delta_j^\pm a) E_{k,\pm j}^{(2)}]_{-{\hat 0\over 2}}
\end{eqnarray}
so that using this approximation, we obtain an approximate solution to Eq.~(\ref{eq:ElecFldEOM}) as
\begin{eqnarray}\label{eq:ElecFldEOMapprox}
\left. E_{i,+{\hat 0\over 2}} \right|_{1} &=& E_{i,-{\hat 0\over 2}} + \Delta t\,[...] +~ \Delta t\,\left({e^2\over 4\pi^2 M}{1\over4}(2+dx\Delta_i^+)\sum\limits_{\pm}\sum\limits_{jk}\epsilon_{ijk}\left[(\Delta_j^\pm a) E_{k,\pm j}^{(2)}\right]_{-{\hat 0\over 2}}\right),\nonumber\\
\end{eqnarray}
The solution for the updated electric field found this way makes Eq.~(\ref{eq:ElecFldApprox}) to reproduce the continuum Eq.~(\ref{eq:EOMaxion2}) to order $\mathcal{O}(dt)$. We can however build now an iterative solution as
\begin{eqnarray}
\label{eq:ElecFldEOMapprox2}\\
\left. E_{i,+{\hat 0\over 2}} \right|_{2} &=& \left. E_{i,+{\hat 0\over 2}}\right|_{1} +~ \Delta t\,\left({e^2\over 4\pi^2 M}{1\over4}(2+dx\Delta_i^+)\sum\limits_{\pm}\sum\limits_{jk}\epsilon_{ijk}\left[(\Delta_j^\pm a) \left. E_{k,\pm j}^{(2)}\right|_{1}\right]_{+{\hat 0\over 2}}\right),\nonumber\\\\
&\vdots&\nonumber\\\label{eq:ElecFldEOMapprox3}
\left. E_{i,+{\hat 0\over 2}} \right|_{n} &=& \left. E_{i,+{\hat 0\over 2}}\right|_{n-1} +~ \Delta t\,\left({e^2\over 4\pi^2 M}{1\over4}(2+dx\Delta_i^+)\sum\limits_{\pm}\sum\limits_{jk}\epsilon_{ijk}\left[(\Delta_j^\pm a) \left. E_{k,\pm j}^{(2)}\right|_{n-1}\right]_{+{\hat 0\over 2}}\right),\nonumber\\
\end{eqnarray}
so that by successive iterations we approach closer an closer to the correct solution to Eq.~(\ref{eq:ElecFldEOM})
\begin{equation}
\left. E_{i,+{\hat 0\over 2}} \right|_{n}\xrightarrow{~n \rightarrow \infty~}E_{i,+{\hat 0\over 2}}
\end{equation}
Of course, it is enough, in principle, to iterate just two times, so that $E_{i,+{\hat 0\over 2}} \simeq \left. E_{i,+{\hat 0\over 2}} \right|_{2}$ solves Eq.~(\ref{eq:ElecFldEOM}) reproducing the continuum Eq.~(\ref{eq:EOMaxion2}) to order $\mathcal{O}(dt^2)$. On the other hand, the Gauss law within the set of Eqs.~(\ref{eq:EOMlatticeAxion}), should be exact (up to computer machine precision) as long as $E_{i}$ is the exact solution to Eq.~(\ref{eq:ElecFldEOM}). However, as we are now approximating the electric field at each time step as $E_{i} \simeq \left. E_{i} \right|_{n}$, it is not clear {\it a priori} the accuracy attained in the (now approximated) Gauss law
\begin{eqnarray}\label{eq:GaussLawApprox}
\sum\limits_i \Delta_i^- \left. E_{i} \right|_{n} &\simeq & {2e^2}{\rm Im}\lbrace\varphi^*\pi\rbrace - {e^2\over 4\pi^2 M}{1\over8}\sum\limits_{\pm}\sum\limits_i (\Delta_i^{\pm}a)(B_i^{(4)}+B_{i,+0}^{(4)})_{\pm i}\,,
\end{eqnarray}
particularly after only $n = 2$ iterations. Even though successive solutions $\left. E_{i} \right|_{n}$ with increasingly larger $n$, can never be better than order $\mathcal{O}(dt^2)$ with respect $\left. E_{i} \right|_{2}$, it might very well be the case that, in order to fulfill the Gauss law with sufficient precision, $\left. E_{i} \right|_{n}$ is required to an order $n \gg 2$. We have not investigated explicitly this aspect in simulations, as this will depend most likely on the specific scenario under study. We leave therefore this check as a future task to be considered when applying our formalism into specific scenarios where a time-dependent and fully-inhomogeneous axion may play a central role.

As a last comment, let us note that in relevant cosmological scenarios like e.g.~axion-inflation~\cite{Freese:1990rb,Sorbo:2011rz,Cook:2011hg,Barnaby:2011qe,Linde:2012bt,Pajer:2013fsa,Adshead:2015pva,Adshead:2016iae}, gauge fields are largely excited due to their axionic-coupling to a shift-symmetric field $a(x)$ which plays the role of the inflaton, and hence is (mostly) homogeneous. Only when the gauge fields are largely excited towards the end of inflation or during preheating, will they back react into the axion field, breaking its (classical) homogeneity. For most of the dynamics the axion-inflaton field remains therefore almost homogeneous. It is therefore conceivable that one may solve the system of Eqs.~(\ref{eq:EOMlatticeAxion}) simply using Eq.~(\ref{eq:ElecFldEOMapprox}) to solve for the electric fields, i.e.~with $E_{i} \simeq \left. E_{i} \right|_{1}$, and yet maintain a good accuracy close to $\mathcal{O}(dt^2)$. The reason for this is that even though the approximated terms in the $rhs$ of Eq.~(\ref{eq:ElecFldEOMapprox}) have a reduced accuracy of $\mathcal{O}(dt)$ instead of $\mathcal{O}(dt^2)$, they are also suppressed by $\vec\nabla\alpha$. Thus, these terms may be negligible in the dynamics, allowing to solve iteratively the set of Eqs.~(\ref{eq:EOMlatticeAxion}) together with Eq.~(\ref{eq:ElecFldEOMapprox}) to advance the electric fields, yet with $\mathcal{O}(dt^2)$ accuracy. As only dedicated simulations can resolve this issue, we leave as future work the test of this circumstance within these scenarios.

\section{Lattice Chern-Simons number(s)}
\label{sec:CSnum}

Let us now consider the definition of the Chern-Simons number in the continuum theory
\begin{eqnarray}\label{eq:CScontinuum}
n_{\rm cs} \equiv {1\over(4\pi)^2}{1\over4}\int d^4x F_{\mn}\tilde{F}^{\mn} = {1\over(4\pi)^2}\int dt \int d^3x ~\vec E \cdot \vec B\,.
\end{eqnarray}
In the lattice, we can define an equivalent quantity describing this topological number, by considering some lattice representation of $\vec E \cdot \vec B$, and substituting the space-time integral by finite sums, $\int d^4x \rightarrow \Delta t \Delta x^3 \sum_{n_o,\vec n}$. In order to do this, we just need to follow a similar logic as in Sect.~\ref{sec:LatticeFormulation}, when we disccused the different lattice representations of an axionic-coupling. For example, we want that the lattice representations of $\vec E$ and $\vec B$ live at the same space-time site, so that we expand the lattice version of $\vec E \cdot \vec B$ around a common site, in order to reproduce the continuum limit to a given order $\mathcal{O}(dx^n)$. Obviously, this prevent us from simply using our ordinary representation of the lattice electric and magnetic fields, $E_i \equiv (\Delta_o^+A_i-\Delta_i^+A_o)$ and $B_i \equiv (\Delta_j^+A_k-\Delta_k^+A_j)$, as these expressions reproduce their continuous analogues to order $\mathcal{O}(dx^2)$ only when interpreting that $E_{i}$ and $B_i$ live at different space-time sites, $l = (n_o+{1\over2},\vec n)$ and $l = (n_o,\vec n+{1\over2}\hat j+{1\over2}\hat k)$, respectively. We could consider to expand these lattice fields around the common space-time lattice site $(n_o,\vec n)$, but then the topological density built like that, $\sum_i E_iB_i = \sum_{i,j,k} \epsilon_{ijk}(\Delta_o^+A_i-\Delta_i^+A_o)\Delta_j^+A_k$, would only reproduce the continuum limit to linear order $\mathcal{O}(dx)$. As we have already derived the lattice EOM reproducing the system continuum dynamics up to order $\mathcal{O}(dx^2)$, we should clearly aim now for a description of the Chern-Simons density to (at least) order $\mathcal{O}(dx^2)$. 

Given our experience in Sect.~\ref{sec:LatticeFormulation} in building up an axionic-type coupling to gauge fields up to order $\mathcal{O}(dx^2)$, a natural candidate to describe the Chern-Simons density at a common space-time site, is naturally given by $\sum_i E_i^{(4)}B_i^{(4)}$, which reproduces the continuum density $\vec E \cdot \vec B$ around $l = (n_o,n)$ to order $\mathcal{O}(dx^2)$. Therefore, we propose as a lattice candidate to describe the Chern-Simons number in the lattice, the following expression
\begin{eqnarray}\label{eq:CSdiscrete1}
n_{\rm cs}^{L(1)} \equiv {1\over(4\pi)^2} \Delta t \Delta x^3 \sum_{n_o,\vec n} \sum_i E_i^{(4)}B_i^{(4)}
\end{eqnarray}
A well known identity in the continuum is
\begin{eqnarray}\label{eq:EBdx4EqualsABdx3Cont}
16\pi^2n_{\rm cs} \equiv \int_0^t dt \int d^3x ~\vec E \cdot \vec B = {1\over2}\int d^3x ~(\vec A \cdot \vec B)(t) ~~+~ \mathcal{C}_o
\end{eqnarray}
where the additive constant is simply given by the initial value 
\begin{equation}
\mathcal{C}_o \equiv - {1\over2}\int d^3x \,(\vec A \cdot \vec B)(0)
\end{equation}
We can easily demonstrate this identity using integration by parts and the fact that fields vanish at infinity: $\int d^4x ~\vec E \cdot \vec B$ = $\int d^4x \,(\dot{A}_i-\partial_i\phi)\epsilon_{ijk}\partial_jA_k$ = $- \int d^4x \,A_k\epsilon_{ijk}\partial_j(\dot{A}_i-\partial_i\phi)$ = $\int d^4x \,A_k\dot{B}_k$ = $\int d^3x \,A_iB_i\big|_{0}^{t} - \int d^4x ~B_k\dot{A}_k$ = $\int d^3x \,A_i B_i\big|_{0}^{t}$ $- \int d^4x \,E_iB_i$ + $\int d^4x \,\epsilon_{ijk}\partial_i\phi\partial_jA_k$, where the last term is equal to $- \int d^4x A_k\epsilon_{ijk}\partial_i\partial_j\phi = 0$, and hence $2\int_0^t\,dt\int d^3x \,\vec E \cdot \vec B = \int d^3x \,\vec A \cdot \vec B\big|_{0}^{t}$. Therefore, a good starting point to check whether Eq.~(\ref{eq:CSdiscrete1}) describes well a Chern-Simons number, would be to see if it verifies the analogous property to Eq.~(\ref{eq:EBdx4EqualsABdx3Cont}) in the lattice. Of course, in the lattice we cannot represent an infinite volume, but rather we take periodic boundary conditions to mimic this. Hence, our demonstration should rely on the use of periodic boundary conditions, which in any case we have implicitly assumed in previous derivations like e.g.~the discrete EOM Eqs.~(\ref{eq:EOMlatticeChemPot}). 

Let us define
\begin{equation}
A_i^{(2)} \equiv {1\over2}(A_i+A_{i,-i}) = {1\over2}(2-dx\Delta_i^-)A_i\,,
\end{equation}
which naturally lives at $l = (n_o,\vec n)$. Let $p,q$ to be non-negative integer numbers. We can now observe that due to the periodic boundary conditions, the following property holds 
\begin{eqnarray}\label{eq:LatticeABtoEBtrick}
\sum_{\vec n} \sum_i A_{i,+p\hat 0}^{(2)}B_{i,+q\hat 0}^{(4)} &=& \sum_{\vec n} \sum_{i,j,k} A_{i,+p\hat 0}^{(2)}\epsilon_{ijk}(\Delta_j^++\Delta_j^-)A_{k,+q\hat 0}^{(2)} \nn &=& \sum_{\vec n} \sum_{i,j,k} \left[\epsilon_{kji}(\Delta_j^++\Delta_j^-)A_{i,+p\hat 0}^{(2)}\right]A_{k,+q\hat 0}^{(2)}\nn
&\equiv& \sum_{\vec n} \sum_i B_{i,+p\hat 0}^{(4)}A_{i,+q\hat 0}^{(2)}
\end{eqnarray}
Essentially, inside a lattice sum $\sum_{\vec n} \sum_i$, one can always exchange $A_{i,+p\hat 0}^{(2)}B_{i,+q\hat 0}^{(4)}$ by $A_{i,+q\hat 0}^{(2)}B_{i,+p\hat 0}^{(4)}$, precisely thanks to the periodicity of the boundary conditions. Thanks to this property, it is easy to check that
\begin{eqnarray}
\sum_{n_o,\vec n} \sum_i E_i^{(4)}B_i^{(4)} &=& {1\over 2\Delta t}\sum_{n_o,\vec n}\sum_{i} (A_{i,+(n_o+1)\hat 0}^{(2)}-A_{i,+(n_o-1)\hat 0}^{(2)})B_i^{(4)} \nn &=& {1\over 2\Delta t}\sum_{n_o,\vec n}\sum_{i} \left(A_{i,+(n_o+1)\hat 0}^{(2)}B_i^{(4)} - A_{i,+(n_o-1)\hat 0}^{(2)}B_i^{(4)}\right) \nn
&\equiv &  {1\over 2\Delta t}\sum_{\vec n}\sum_{i}A_{i,+(p-1)\hat 0}^{(2)}B_{i,+p \hat 0}^{(4)} ~~~+~ \mathcal{D}_{o}\,,
\end{eqnarray}
where the sum over time steps has been explicitly divided in $p$ steps, and the additive constant is given by the fields' initial value $\mathcal{D}_{o} \equiv - {1\over 2\Delta t}\sum_{n}\sum_{i}A_{i,-\hat 0}^{(2)}B_{i}^{(4)}$. We have just demonstrated the following lattice identity
\begin{eqnarray}\label{eq:EBdx4EqualsABdx3Discrete}
16\pi^2 n_{\rm cs}^{L(1)} &=& \Delta t \Delta x^3 \sum_{n_o = 0}^p\sum_{\vec n} \sum_i E_i^{(4)}B_i^{(4)} \nn
&=& {1\over 2}\Delta x^3 \sum_{\vec n} \sum_i A_{i,+(p-1)\hat 0}^{(2)}B_{i,+p \hat 0}^{(4)} ~~~+~ \mathcal{D}_o^{(1)}\,,
\end{eqnarray}
with
\begin{equation}
\mathcal{D}_o^{(1)} = - {1\over 2}\Delta x^3\sum_{\vec n}\sum_{i}A_{i,-\hat 0}^{(2)}B_{i}^{(4)}\,.
\end{equation}

Our lattice expression for the Chern-Simons number Eq.~(\ref{eq:CSdiscrete1}) successfully passes the first check, as it verifies the identity Eq.~(\ref{eq:EBdx4EqualsABdx3Discrete}), which clearly represents the analogue to the continuum identity Eq.~(\ref{eq:EBdx4EqualsABdx3Cont}). 

\subsection{Chern-Simons number and chemical potential}

Let us consider now an Abelian gauge theory with chemical potential $\mu$. As a consequence of integrating in time the EOM for the chemical potential Eq.~(\ref{eq:EOMChPot4}), and using the Chern-Simons number definition Eq.~(\ref{eq:CScontinuum}), the following relation follows in the continuum
\begin{equation}\label{eq:MuCSrelationContinuum}
\mu(t) = \mu(0) + {48\over T^2}\lim_{V\rightarrow \infty}{n_{\rm cs}\over V}\,.
\end{equation}
Therefore, a relevant test we can impose over our lattice implementation of the Chern-Simons number Eq.~(\ref{eq:EBdx4EqualsABdx3Discrete}), is to see whether it verifies some lattice version of the relation Eq.~(\ref{eq:MuCSrelationContinuum}). From the discrete evolution equation for the chemical potential, see Eqs.~(\ref{eq:EOMlatticeChemPot}), %reads $\Delta_o^+\mu = {3\over \pi^2} {1\over T^2}{1\over 2N^3}\sum_{\vec n} \sum_i E_i^{(2)}(B_i^{(4)}+B_{i,+0}^{(4)})$. From here, 
we see that the chemical potential after $p$ time steps, is given by
\begin{eqnarray}\label{eq:MuCSrelationDiscrete}
\mu_{+p\hat 0} &=& \mu_o + {3\over \pi^2} {1\over T^2V_L^3}{1\over 2}\Delta t \Delta x^3\sum_{n_o = 0}^{p-1}\sum_{\vec n} \sum_i E_i^{(2)}(B_i^{(4)}+B_{i,+0}^{(4)})\nonumber\\
&\equiv& \mu_o + {48\over T^2V_L^3}n_{\rm cs}^{L(2)}\,,
\end{eqnarray}
where we denote the initial chemical potential value as $\mu_o \equiv \mu(n_o = 0)$, and the lattice volume as $V_L \equiv (Ndx)^3$, with $N$ is the number of lattice sites per dimension. Eq.~(\ref{eq:MuCSrelationDiscrete}) has lead us in fact to define a new lattice Chern-Simons number as
\begin{eqnarray}%\label{eq:CSdiscrete2}
\label{eq:EBdx4EqualsABdx3Discrete2}
16\pi^2 n_{\rm cs}^{L(2)} &\equiv& {\Delta t \Delta x^3}\sum_{n_o=0}^{p-1} \sum_{\vec n} {1\over 2}\sum_i E_i^{(2)}(B_i^{(4)}+B_{i,+0}^{(4)}) \nn
&=& {\Delta x^3\over 2}\sum_{n_o=0}^{p-1} \sum_{\vec n} \sum_i \left\lbrace A_{i,+0}^{(2)}B_{i}^{(4)} + A_{i,+0}^{(2)}B_{i,+0}^{(4)} - A_{i}^{(2)}B_{i}^{(4)} - A_{i}^{(2)}B_{i,+0}^{(4)}\right\rbrace \nn
&=& {\Delta x^3\over 2}\sum_{\vec n} \sum_i A_{i,+p\hat 0}^{(2)}B_{i,+p \hat 0}^{(4)} ~~~+~ \mathcal{D}_o^{(2)}\,,%\label{eq:EBdx4EqualsABdx3Discrete2}
\end{eqnarray}
with initial constant
\begin{equation}\label{eq:C2}
~~~ \mathcal{D}_o^{(2)} = - {1\over 2}\Delta x^3\sum_{\vec n}\sum_{i}A_{i}^{(2)}B_{i}^{(4)}\,.
\end{equation}
Notice that in order to arrive at the final expression of Eq.~(\ref{eq:EBdx4EqualsABdx3Discrete2}) of the form $\sim \sum_{\vec n} AB$ from the initial expression of the form $\sim \sum_{n_o,\vec n} EB$, we have applied again the trick expressed by Eq.~(\ref{eq:LatticeABtoEBtrick}). We conclude that in order to correctly mimic the continuum relation Eq.~(\ref{eq:MuCSrelationContinuum}), we are forced to define a second Chern-Simons number by Eq.~(\ref{eq:EBdx4EqualsABdx3Discrete2}), which is similar to that of Eq.~(\ref{eq:EBdx4EqualsABdx3Discrete}), but still different. As a matter of fact, we have already shown in Eq.~(\ref{eq:EBdx4EqualsABdx3Discrete2}) that this new Chern-Simons verifies a lattice analogue to the continuum relation Eq.~(\ref{eq:EBdx4EqualsABdx3Cont}) $\int dx^4 ~\vec E\cdot \vec B = {1\over2}\int dx^3 \vec A\cdot \vec B ~+~ \mathcal{C}_o$. A simple check also shows that $n_{\rm cs}^{L(2)}$ is built from a Chern-Simons lattice density which reproduces exactly the continuum expression $\vec E\cdot \vec B$ up to order $\mathcal{O}(dx^2)$. 

A Chern-Simons number given by Eq.~(\ref{eq:EBdx4EqualsABdx3Discrete2}) is therefore, not only equally valid as Eq.~(\ref{eq:EBdx4EqualsABdx3Discrete}), but also preferable. Contrary to Eq.~(\ref{eq:EBdx4EqualsABdx3Discrete}), it is exactly linearly proportional to the chemical potential at arbitrary times, see Eq.~(\ref{eq:MuCSrelationDiscrete}), as expected in order to mimic the continuum relation Eq.~(\ref{eq:MuCSrelationContinuum}). Besides, as our final choice for the axion-gauge interaction is based on the lagrangian density $E_i^{(2)}(B_i^{(4)}+B_{i,+0}^{(4)})$, see Eq.~(\ref{eq:LatticeaxionActionFinal}), there is no surprise whatsoever that such expression [and not $E_i^{(4)}B_i^{(4)}$ on which Eq.~(\ref{eq:EBdx4EqualsABdx3Discrete}) is based on] defines the natural Chern-Simons number of the system: it is such density the one naturally sourcing the discrete EOM of the chemical potential, see Eqs.~(\ref{eq:EOMlatticeChemPot}).

{\it Does this mean that our expression $n_{\rm cs}^{L(1)}$ Eq.~(\ref{eq:CSdiscrete1}) is actually incorrect ?} Actually not really, as the different between $n_{\rm cs}^{L(2)}$ Eq.~(\ref{eq:EBdx4EqualsABdx3Discrete2}) and $n_{\rm cs}^{L(1)}$ Eq.~(\ref{eq:CSdiscrete1}) can be exactly determined. With a bit of algebra, we find
\begin{eqnarray}\label{eq:CS1vsCS2}
&& {(4\pi)^2\over V_L}\left(n_{\rm cs}^{L(2)} - n_{\rm cs}^{L(1)}\right) \nn
&& \equiv {1\over N^3}\sum_{\vec n} \sum_i \left\lbrace\left(A_{i,+p\hat 0}^{(2)}-A_{i,+(p-1)\hat 0}^{(2)}\right)B_{i,+p \hat 0}^{(4)} - dt E_{i}^{(2)}B_i^{(4)} \right\rbrace \sim \mathcal{O}(dt)\,,
\end{eqnarray}
Note that we have used the fact that the $\left(A_{i,+p\hat 0}^{(2)}-A_{i,+(p-1)\hat 0}^{(2)}\right)$ is simply\footnote{By construction $\left(A_{i,+p\hat 0}^{(2)}-A_{i,+(p-1)\hat 0}^{(2)}\right) \equiv dtE_{i}((p-{1\over2})\hat 0,\vec n+{1\over2}\hat i)$, so even though it changes in time, this change is not cumulative, and represents always a small perturbation.} of order $\mathcal{O}(dt)$, whereas the initial constant $-dt E_{i}^{(2)}B_i^{(4)}$ is, of course, of order $\mathcal{O}(dt)$ by construction. Therefore, we see that both lattice Chern-Simon numbers 'track' each other in time, and their relative (density) difference is always a small $\mathcal{O}(dt)$ perturbation. We have in fact numerically checked the relation Eq.~(\ref{eq:CS1vsCS2}) in lattice simulations, and found that it is verified exactly (to machine precision). In conclusion, even though both lattice Chern-Simons numbers represent a valid description, given that $n_{\rm cs}^{L(2)}$ Eq.~(\ref{eq:EBdx4EqualsABdx3Discrete2}) sources exactly the chemical potential [whereas $n_{\rm cs}^{L(1)}$ Eq.~(\ref{eq:EBdx4EqualsABdx3Discrete}) does not], we propose to work only with $n_{\rm cs}^{L(2)}$. 

Let us also note that using the trick expressed by Eq.~(\ref{eq:LatticeABtoEBtrick}), we can also write the lattice axionic-interaction term in the case of an Abelian gauge theory with chemical potential, in an alternative way to Eq.~(\ref{eq:LatticeHomAxionAction}), like
\begin{eqnarray}\label{eq:LatticeaxionActionAB}
S_{\alpha(t)}^{L} &=& \Delta t \Delta x^3\sum_{\vec n,n_o}\left\lbrace {M^2\over 2c_s^2}\left(\Delta_o^-\alpha\right)^2 + {1\over 4\pi^2}{\alpha\over2}\sum_i E_i^{(2)}\left(B_i^{(4)}+B_{i,+0}^{(4)}\right)\right\rbrace \nn
&\equiv & \Delta t \Delta x^3\sum_{n_o}\left\lbrace N^3{M^2\over 2c_s^2}\left(\Delta_o^-\alpha\right)^2 - {1\over 4\pi^2}(\Delta_o^-\alpha){1\over2}\sum_{\vec n} \vec{A}_i^{(2)}\vec{B}^{(4)}\right\rbrace\,,
%\\&=& {e^2\over 4\pi^2}\sum_{\vec n,n_o} {\alpha(l)\over16} \sum_{i,j,k} \epsilon_{ijk}[(2-\delta_i^-)(\Delta_o^+ A_i)](l)[(\Delta_j^+ + \Delta_j^-)(2-\delta_k^-)(2+\delta_o^+)A_k](l)\Big|_{l \equiv (n,n_o+\hat 0/2)}\,,
\end{eqnarray}
where %we have used $\mu \equiv \Delta_o^- \alpha$ and, 
we have discarded additive constants in the last expression of Eq.~(\ref{eq:LatticeaxionActionAB}), as they do no contribute to the EOM when varying the action. One can check, of course, that from variations of the last expression in Eq.~(\ref{eq:LatticeaxionActionAB}), we also obtain identical EOM as in Eq.~(\ref{eq:EOMlatticeChemPot}), as it should be. Writing $S_{\alpha(t)}^{L}$ as in the last expression in Eq.~(\ref{eq:LatticeaxionActionAB}), is perhaps the most natural thing to do in the presence of a chemical potential (homogeneous axion), as $\mu \equiv \Delta_o^{-}\alpha$ is simply given by the time derivative of an auxiliar variable $\alpha$, but the latter plays no dynamical role. The last expression in Eq.~(\ref{eq:LatticeaxionActionAB}) eliminates precisely the presence of $\alpha$ in the action, and maintains only terms involving explicitly $\mu = \Delta_o^{-}\alpha$.

\subsection{Derivative representation of Pontryagin density $\tilde{F}_\mn F^\mn$}

Our former demonstration(s) that the Chern-Simons number(s) can be written in either form $\sim \sum_{n_o,\vec n} \vec E \vec B$ or as $\sim \sum_{\vec n} \vec A \vec B$, indicates that we can find a derivative representation of our lattice expressions of the Pontragyn density. Whereas in the continuum we can write the identity $Q \equiv \tilde{F}_\mn F^\mn = \partial_\mu K^\mu$, with $K^\mu$ the Chern-Simons current, we expect that in the discrete some analogous relation may exist, so that $Q = \Delta_\mu^+ K^\mu$. In order to see this, let us focus on the Chern-Simons number $n_{\rm cs}^{L(2)}$ (analogous derivations can be applied to $n_{\rm cs}^{L(1)}$). If we redo the steps in Eq.~(\ref{eq:EBdx4EqualsABdx3Discrete2}), we immediately realize that the lattice Chern-Simons number $n_{\rm cs}^{L(2)}$ can be re-written like
\begin{eqnarray}%\label{eq:CSdiscrete2}
\label{eq:K0intro}
16\pi^2 n_{\rm cs}^{L(2)} &\equiv& {\Delta t \Delta x^3}\sum_{n_o=0}^{p-1} \sum_{\vec n} Q_L = {\Delta t \Delta x^3}\sum_{n_o=0}^{p-1}\sum_{\vec n} \Delta_o^+K^o
\end{eqnarray}
where
\begin{eqnarray}\label{eq:Qlattice}
Q_L \equiv {1\over 2}\sum_i E_i^{(2)}(B_i^{(4)}+B_{i,+0}^{(4)})\,,
\end{eqnarray}
and
\begin{eqnarray}\label{eq:K0def}
K^0 = - K_0 \equiv {1\over 2}\sum_i A_{i}^{(2)}B_{i}^{(4)}
\end{eqnarray}
Note that this does not mean that we can locally substitute the expression for the lattice Pontryagin density $Q_L$ by $\Delta_o^+K^o$, like if it was an identity at every lattice site.  However, whenever summing over the lattice volume, we can do such a replacement locally inside the argument of the lattice sum, i.e.~
\begin{eqnarray}\label{eq:QtoK0}
\sum_{\vec n} Q_L = \sum_{\vec n}\Delta_o^+K^o\,.
\end{eqnarray}
If we also sum over the time history of the system, we then arrive immediately at
\begin{eqnarray}%\label{eq:CSdiscrete2}
\label{eq:K0check}
16\pi^2 n_{\rm cs}^{L(2)} &\equiv& {\Delta t \Delta x^3}\sum_{n_o=0}^{p-1} \sum_{\vec n} Q_L = {\Delta t \Delta x^3}\sum_{n_o,\vec n} \Delta_o^+K^o = \Delta x^3\left(\sum_{\vec n}K^o_{+p\hat 0} - \sum_{\vec n}K^o\right)\,.\nn
\end{eqnarray}
Using the definition of $K^o$ in Eq.~(\ref{eq:K0def}), we see that Eq.~(\ref{eq:K0check}) coincides exactly, as it should, with the final expression of Eq.~(\ref{eq:EBdx4EqualsABdx3Discrete2}).

Let us find the spatial $K^i$ components of the lattice representation of the Chern-Simons current. Starting from the definition of $Q_L$ in Eq.~(\ref{eq:Qlattice}), we can write
\begin{eqnarray}\label{eq:QgoingDK}
Q_L &\equiv& {1\over 2}\sum_i \Delta_o^+A_i^{(2)}(B_i^{(4)}+B_{i,+0}^{(4)})\nn
&\equiv& {1\over\Delta t}\sum_i \left\lbrace{1\over2}\left(A_{i,+0}^{(2)}B_{i,+0}^{(4)}-A_i^{(2)}B_i^{(4)}\right) + {1\over2}\left(A_{i,+0}^{(2)}B_{i}^{(4)}-A_i^{(2)}B_{i,+0}^{(4)}\right) \right\rbrace\,,
\end{eqnarray}
and immediately recognize the first term in the $rhs$ of Eq.~(\ref{eq:QgoingDK}) as equal to $\Delta_0^+K^0$. On the other hand, the second term in the $rhs$ of Eq.~(\ref{eq:QgoingDK}), can be re-written like
\begin{eqnarray}\label{eq:QgoingDK2}
&& {1\over2\Delta t}\sum_i\left(A_{i,+0}^{(2)}B_{i}^{(4)}-A_i^{(2)}B_{i,+0}^{(4)}\right) \nn
&=& {1\over4\Delta t\Delta x}\sum_{i,j,k} \left\lbrace A_{i,+0}^{(2)}\epsilon_{ijk}\left(A_{k,+j}^{(2)}-A_{k,-j}^{(2)}\right)-A_i^{(2)}\epsilon_{ijk}\left(A_{k,+\hat 0 + \hat j}^{(2)} - A_{k,+\hat 0 - \hat j}^{(2)}\right)\right\rbrace\nn
&=& {1\over4\Delta x}\sum_{i,j,k}\left\lbrace \left(E_{i,+j}^{(2)}\epsilon_{ijk}A_{k}^{(2)} + E_{i}^{(2)}\epsilon_{ijk}A_{k,+j}^{(2)}\right) - \left(E_i^{(2)}\epsilon_{ijk}A_{k,-j}^{(2)} + E_{i,-j}^{(2)}\epsilon_{ijk}A_{k}^{(2)}\right)\right\rbrace\nn
&\equiv& \sum_j\Delta_j^+K^j\,,
\end{eqnarray}
where we have identified
\begin{eqnarray}\label{eq:KiDef}
K^i = K_i \equiv -{1\over4}\sum_{j,k}\epsilon_{ijk}\left(E_{j}^{(2)}A_{k,-i}^{(2)}+E_{j,-i}^{(2)}A_{k}^{(2)}\right)\,.
\end{eqnarray}

We have demonstrated therefore that we can write 
\begin{eqnarray}\label{eq:QequalDK}
Q_L = \sum_{\mu}\Delta_\mu^+ K^\mu = \Delta_0^+ K^0 + \Delta_1^+ K^1 + \Delta_2^+ K^2 + \Delta_3^+ K^3\,,
\end{eqnarray}
as a local identity at every lattice site, with $K^\mu$ defined by components, $K^0$ given by Eq.~(\ref{eq:K0def}) and $K^i$ given by Eq.~(\ref{eq:KiDef}). Given that we consider periodic boundary conditions, let us note that it must be true that $\sum_{\vec n} \sum_{i}\Delta_i^+K^i = 0$, no matter the expression for $K^i$. Therefore, the identity in Eq.~(\ref{eq:QtoK0}) will not change in any case, if we just substitute $\Delta_o^+K^o$ by $\sum_{\mu}\Delta_\mu^+K^\mu$ in the $rhs$ of such equation,
\begin{eqnarray}
\sum_{\vec n} Q_L = \sum_{\vec n}\sum_{\mu}\Delta_\mu^+K^\mu = \sum_{\vec n}\Delta_o^+K^o\,.
\end{eqnarray}
This implies that the expression for the Chern-Simons number Eq.~(\ref{eq:K0check}), given only in terms of $K^0$ (and not $K^i$'s), is of course unchanged,
\begin{eqnarray}
16\pi^2 n_{\rm cs}^{L(2)} = {\Delta t \Delta x^3}\sum_{n_o,\vec n}\sum_\mu \Delta_\mu^+K^\mu = {\Delta t \Delta x^3}\sum_{n_o,\vec n} \Delta_0^+K^0 = \Delta x^3\left(\sum_{\vec n}K^o_{+p\hat 0} - \sum_{\vec n}K^o\right)\nn
\end{eqnarray}
as it should, independently of the expression Eq.~(\ref{eq:KiDef}) we found for $K^i$.

We have found, as promised, a lattice expression for the Pontryagin density, Eq.~(\ref{eq:QequalDK}), that admits a total derivative representation, with $K^0$ and $K^i$ given by Eqs.~(\ref{eq:K0def}), (\ref{eq:KiDef}), respectively.

\subsection{Chern-Simons number in the presence of a magnetic field}

Let us now turn our attention into the case where a background magnetic field is present in the system. Following~\cite{Kajantie:1998rz}, we can introduce a magnetic flux in the lattice by demanding that the boundary conditions of the gauge fields $A_i$ are not periodic. Without loss of generality, we can consider an external magnetic field in the $\hat z$ direction, $\vec B = (0,0,B)$. Such magnetic field can be introduced in the lattice, by demanding that only the component $A_1(n_1,n_2,n_3)$ of the gauge field is 'aperiodic' at a given $x$-site, say $n_1 = 1$, at the boundary of the lattice $y$-axis, independently of its location within the $z$-axis, i.e.~
\begin{equation}\label{eq:twisted}
dx\left(A_j(n_1,0,n_3) - A_j(n_1,N,n_z)\right) = 2\pi n_{\rm mag}\delta_{1j}\delta_{1n_1}\,.
\end{equation}
This condition creates a flux of magnitude $\Phi_{\rm flux} =\equiv 2\pi n_{\rm mag}$ orthogonal to the $xy$-plane of the lattice, with area $A \equiv (Ndx)^2$,
\begin{eqnarray}\label{eq:Bmag}
\int_A \vec B \,d^2\vec x = \int_A B \,dx_1dx_2 = B (Ndx)^2 \equiv 2\pi n_{\rm mag}\,,~~~\Rightarrow~~~ B \equiv {2\pi n_{\rm mag}\over (dx N)^2}\,.
\end{eqnarray}
In principle, any flux can be generated. However, quantization of this flux is required for maintaining a periodic action in the non-compact formulation of a gauge theory, see~\cite{Kajantie:1998rz} for details. We need therefore to take $n_{\rm mag}$ as a positive integer, with $n_{\rm mag} = 0$ simply representing the absence of magnetic field. 

The 'twisted' boundary condition for $A_1$ Eq.~(\ref{eq:twisted}) implies that whenever we want to calculate a magnetic field $B_3(n_1,n_2,n_3)$ at the location $(n_1,n_2,n_3) = (1,N-1,n_3)$, instead of simply computing $(\Delta_1^+ A_2 - \Delta_2^+ A_1)$, we should really make the calculation
\begin{equation}
B_3(n_1,n_2,n_3) ~~~~\longrightarrow~~~~ (\Delta_1^+ A_2 - \Delta_2^+ A_1) - {2\pi n_{\rm mag}\over dx^2}\delta_{1n_1}\delta_{(N-1)n_2}\,,
\end{equation}
as this is equivalent to assume that the first component of the gauge field makes a 'discrete jump' $A_1(1,0,n_3) = A_1(1,N,n_3) + {2\pi n_{\rm mag}\over dx}$, just as required by Eq.~(\ref{eq:twisted}).

Let us now see how the Chern-Simons number Eq.~(\ref{eq:EBdx4EqualsABdx3Discrete2}) is modified in the presence of a background magnetic field, introduced in the lattice through the condition Eq.~(\ref{eq:twisted}). Let us begin by noticing that the relation Eq.~(\ref{eq:LatticeABtoEBtrick}), used previously to relate the $\sim EB$ and $\sim AB$ expressions of the CS number [see e.g.~Eq.~(\ref{eq:EBdx4EqualsABdx3Discrete}) or Eq.~(\ref{eq:EBdx4EqualsABdx3Discrete2})], does not hold anymore, as it relies on the (now lost) periodic boundary conditions of the gauge fields. When the gauge field follows instead the twisted boundary condition Eq.~(\ref{eq:twisted}), Eq.~(\ref{eq:LatticeABtoEBtrick}) needs to be modified. Given some $p,q$ integer numbers, we obtain now
\begin{eqnarray}\label{eq:LatticeABtoEBtrickII}
&& \sum_{\vec n} \sum_i A_{i,+p\hat 0}^{(2)}B_{i,+q\hat 0}^{(4)} \\
&& = \sum_{\vec n} \sum_{i,j,k} A_{i,+p\hat 0}^{(2)}\left[\epsilon_{ijk}(\Delta_j^++\Delta_j^-)A_{k,+q\hat 0}^{(2)} - {2\pi n_{\rm mag}\over \Delta x^2}\delta_{i3}(\delta_{1n_1}+\delta_{2n_1})(\delta_{0n_2}+\delta_{(N-1)n_2})\right] \nn 
&& = \sum_{\vec n}\left\lbrace\sum_{i,j,k} \left[\epsilon_{kji}(\Delta_j^++\Delta_j^-)A_{i,+p\hat 0}^{(2)}\right]A_{k,+q\hat 0}^{(2)} - {2\pi n_{\rm mag}\over \Delta x^2}A_{3,+p\hat 0}^{(2)}(\delta_{1n_1}+\delta_{2n_1})(\delta_{0n_2}+\delta_{(N-1)n_2})\right\rbrace\nn
&& \equiv \sum_{\vec n} \sum_i B_{i,+p\hat 0}^{(4)}A_{i,+q\hat 0}^{(2)} ~~~+~ \mathcal{M}_{p}\,,
\end{eqnarray}
where $\mathcal{M}_{p}$ represents a magnetic correction given by
\begin{eqnarray}\label{eq:MagConst}
\mathcal{M}_{p} &\equiv& - {2\pi n_{\rm mag}\over \Delta x^2}\sum_{n_3} \lb A_{3,+p\hat 0}^{(2)}(1,0,n_3)+A_{3,+p\hat 0}^{(2)}(1,N-1,n_3) \right. \nn && \left. \hspace*{3cm} + A_{3,+p\hat 0}^{(2)}(2,0,n_3)+A_{3,+p\hat 0}^{(2)}(2,N-1,n_3) \rb\,.
\end{eqnarray}
Essentially, inside a lattice sum $\sum_{\vec n} \sum_i$, one can still substitute $A_{i,+p\hat 0}^{(2)}B_{i,+q\hat 0}^{(4)}$ by $A_{i,+q\hat 0}^{(2)}B_{i,+p\hat 0}^{(4)}$, as long as outside the sum one compensates by adding the magnetic correction $\mathcal{M}_{p}$. This correction is precisely a manifestation of the loss of periodic boundary conditions Eq.~(\ref{eq:twisted}) for the gauge field component $A_1$. 

In light of Eq.~(\ref{eq:LatticeABtoEBtrickII}), the expression of the lattice Chern-Simons number Eq.~(\ref{eq:EBdx4EqualsABdx3Discrete2}), changes in the presence of an external magnetic field to
\begin{eqnarray}\label{eq:EBdx4EqualsABdx3Discrete3}
16\pi^2 n_{\rm cs}^{L(3)} &\equiv& {\Delta t \Delta x^3}\sum_{n_o=0}^p \sum_{\vec n} {1\over 2}\sum_i E_i^{(2)}(B_i^{(4)}+B_{i,+0}^{(4)}) \nn
&=& {\Delta x^3\over 2} \sum_{\vec n} \sum_i A_{i,+p\hat 0}^{(2)}B_{i,+p \hat 0}^{(4)} ~~~+~ \mathcal{D}_o^{(2)} + \mathcal{M}_p - \mathcal{M}_0\,, 
\end{eqnarray}
with $\mathcal{D}_o^{(2)}$ the same constant as in Eq.~(\ref{eq:C2}), and $\mathcal{M}_0, \mathcal{M}_p$ given by Eq.~(\ref{eq:MagConst}), representing respectively initial and final (after $p$ time steps) magnetic field corrections. Equivalently,
\begin{eqnarray}
16\pi^2 n_{\rm cs}^{L(3)} = 16\pi^2 n_{\rm cs}^{L(2)} + \mathcal{M}_p - \mathcal{M}_0\,.
\end{eqnarray}

\section{Summary and Discussion}
\label{sec:Conclusions}

\indent In this paper we have derived a lattice representation of an axionic-interaction $a(x)\tilde{F}_\mn F^\mn$, presenting step by step the necessary ingredients to achieve a formulation that $i)$ reproduces the continuum limit to order $\mathcal{O}(dx^2)$, $ii)$ it is consistent with the (lattice version of the) Bianchi identities, and $iii)$ it is solvable by an iterative scheme of evolution. We first considered in Sect.~\ref{subsec:LatticeHomAxion} the case of a homogeneous axion $a(x) = a(t)$, deriving a lattice representation of the axion-gauge interaction that leads to a set of discrete couple equations solvable by an explicit local iterative scheme, see Eqs.~(\ref{eq:EOMlatticeChemPot}). We generalized our results afterwards, in Sect.~\ref{subsec:LatticeInHomAxion}, to the case of a fully inhomogeneous axion $a(x) = a(t,{\bf x})$. We showed that the set of discrete lattice Eqs.~(\ref{eq:EOMlatticeAxion}) do not admit a simple local explicit solution (while preserving the $\mathcal{O}(dx^2)$ difference with respect to the continuum). We have proposed an implicit scheme to overcome this difficulty. We have also introduced consistent lattice formulation(s) of the Chern-Simons number $n_{\rm cs} \propto \int d^4x\,Q$ (Sect.~\ref{sec:CSnum}) based on the lattice version(s) of $Q = \tilde{F}_\mn F^\mn$ developed in Sect.~\ref{sec:LatticeFormulation}. We put special care in the need to achieve a lattice formulation that admits a total derivative representation for $Q = \Delta_\mu^+K^\mu$. We showed explicitly that such total derivative representation exists, and provided the expression for the $K^\mu$ components, see Eqs.~(\ref{eq:K0def}), (\ref{eq:KiDef}). Finally, we derived the analogous lattice expressions for the Chern-Simons number in the presence of an external magnetic field.

A number of potential applications of our lattice formalism has been already mentioned in the Introduction.  In particular, in an accompanying paper~\cite{Figueroa:2017tbf} we study a number of questions one can address in high temperature electrodynamics with non-zero background magnetic field and fermion chemical potential. This theory was formulated on the lattice in Section~\ref{subsec:LatticeHomAxion}.  We investigate in~\cite{Figueroa:2017tbf}  the random walk of the topological charge and show that it has a diffusive behavior in the presence of magnetic field $B$. This indicates that the mechanism for fermionic number non-conservation for $B\neq 0$ is similar to that in non-Abelian gauge theories. The diffusion rate is related to the rate of chiral charge non-conservation, and we present new results concerning the value  of this rate. Our formulation allows us to study the dynamics of instabilities in the presence of non-zero $\mu$ and elucidate the role of thermal fluctuations of the gauge fields. We find several interesting behaviors that we did not expect $a~priori$.

In addition, let us note that our formalism can be useful as well for the study of the non-linear dynamics in axion-inflation models~\cite{Freese:1990rb,Sorbo:2011rz,Cook:2011hg,Barnaby:2011qe,Linde:2012bt,Pajer:2013fsa}. In these scenarios, gauge fields coupled to a pseudo-scalar inflaton, are excited to high occupation states. Towards the last stages of inflation the system becomes non-linear: the gauge fields are so largely excited, that they significantly back-react into the inflaton dynamics, affecting also inflationary expansion rate. The large amplification of the gauge fields leads to a very efficient generation of gravitational waves and scalar density perturbations, both with non-Gaussian statistics. If the amplitude of the scalar perturbations sourced by the gauge fields is too large, primordial black holes (PBHs) in excess to the current bounds may appear~\cite{Linde:2012bt}. The details of both the gravitational waves and scalar perturbations (possibly leading to PBHs), depend very sensitively on the late non-linear stages of inflation, where analytical techniques can only provide an order of magnitude estimation of the dynamics. Our formalism, however, is suitable for solving the complicate non-linear dynamics numerically on a lattice, deviating from the continuum (classical) theory only to order $\mathcal{O}(dx^2)$. Besides, in axion-inflation scenarios, preheating is driven by the so called {\it tachyonic resonance} of the gauge fields, which occurs precisely due to the axionic-coupling with the inflaton~\cite{Adshead:2015pva,Adshead:2016iae}. This represents a complicate non-linear evolution stage after inflation, for which our lattice formalism can be particularly suitable.

\section*{Acknowledgements}

We thank Dmitry Levkov and Alberto Ramos for clarifications about iterative schemes of evolution to solve implicit chain of discrete equations of motion. This work was supported by the ERC-AdG-2015 grant 694896.  The work of  M.S. was supported partially by the Swiss National Science Foundation. 

\bibliographystyle{utphys}
\bibliography{Lattice}{}

\end{document}